%% file: main_reports_correlations_260718.tex
\documentclass[12pt]{iopart}

\usepackage{graphicx}
\usepackage{color}
\usepackage{hyperref}
\usepackage{bbm}
\usepackage{amssymb}

\newcommand{\eqref}[1]{(\ref{#1})}
\newcommand{\ket}[1]{\left\vert#1\right\rangle}
\newcommand{\bra}[1]{\left\langle#1\right\vert}
\newcommand{\braket}[2]{\left\langle#1|#2\right\rangle}
\newcommand{\id}{\mathbbm{1}}

\begin{document}

\title{Genuine quantum correlations in quantum many-body systems: a review of recent progress}
\author{Gabriele De Chiara$^1$}
\address{$^1$Centre for Theoretical Atomic, Molecular and Optical Physics,
Queen's University Belfast, Belfast BT7 1NN, United Kingdom}
\author{Anna Sanpera$^{2,3}$}
\address{$^2$ICREA, Pg. Llu\'is Companys 23, E-08010 Barcelona, Spain}
\address{$^3$F\'{\i}sica Te\`{o}rica: Informaci\'{o} i Fen\`{o}mens Qu\`{a}ntics, Departament de F\'{\i}sica, Universitat Aut\`{o}noma de Barcelona, 08193 Bellaterra, Spain}

\begin{abstract}
Quantum information theory has considerably helped in the understanding of quantum many-body systems. The role of quantum correlations and in particular, bipartite entanglement, has become crucial to characterise, classify and simulate quantum many body systems. Furthermore, the scaling of entanglement has inspired modifications to numerical techniques for the simulation of many-body systems leading to the, now established, area of tensor networks. However, the notions and methods brought by quantum information do not end with bipartite entanglement. There are other forms of  correlations embedded in the ground, excited and thermal states of quantum many-body systems that also need to be explored and might be utilised as potential resources for quantum technologies.  The aim of this work is to review the most recent developments regarding correlations in quantum many-body systems focussing on multipartite entanglement, quantum nonlocality, quantum discord, mutual information but also other non classical measures of correlations based on quantum coherence.
Moreover, we also discuss applications of quantum metrology in quantum many-body systems.
\end{abstract}
\tableofcontents{}


\input{Chapter1.tex}

\input{Chapter2.tex}

\input{Chapter3.tex}

\input{Chapter4.tex}

\input{Chapter5.tex}

\input{Chapter6.tex}

\ack We acknowledge fruitful discussions with our collaborators on the topics discussed in this review: G. Adesso, V. Ahufinger, T. Apollaro, S. Campbell, L. Correa, R. Fazio, L. Lepori, M. Lewenstein, M. Mehboudi, M. Moreno-Cardoner, S. Paganelli, M. Paternostro, B. Rogers, T. Roscilde and J. Stasinska. 
AS acknowledges financial support 
from the Spanish MINECO FIS2016-80681-P (AEI/FEDER, UE), Generalitat de Catalunya CIRIT  2017-SGR-1127 .
This work was partially done at the Pere Pascual Benasque Center of Sciences (Spain).

\section*{References}
\bibliographystyle{iopart-num}
\bibliography{biblio_all_26072018}

\end{document}

%% file: Chapter1.tex


\section{Introduction}

The study of quantum many-body systems, almost as old as quantum theory itself, has witnessed tremendous progress fostered by impressive advances in the manufacturing of novel materials, their accurate probing at the quantum level and their characterisation with sophisticated numerical simulations. Besides the quest for the creation of exotic materials in solid state systems exhibiting quantum features such as spin liquids \cite{SavaryBalents} and topological insulators \cite{Ando}, the simulation of quantum many-body models by means of quantum simulators, i.e., by extremely well controlled quantum systems, following Feynman's idea, is currently opening the exploration of new quantum phenomena with unprecedented control~\cite{GeorgescuRMP2014,Zohar2016}. Experiments with tens of ultracold atoms in optical lattices and with trapped ions are entering the strongly-correlated regime that is believed to be hard to simulate with classical computers \cite{LewensteinOxford2012}. 
From a fundamental and theoretical perspective a special role is reserved to quantum phase transitions \cite{Sachdev} at which a collective rearrangement of the particle properties and their quantum correlations take place.

Quantum information theory has helped developing a precise language with an
ample vocabulary for the characterisation of quantum systems based on quantum correlations which can be applied to the study of many-body systems often represented e.g. as spin models. Undoubtedly, the theory of entanglement has provided a new set of powerful tools to gain understanding about many-body systems by explicitly considering the tensorial product structure of the composite states. Quantum correlations put in evidence the importance of such tensorial structure. Entangled states of composite systems are those which do not admit a description of them in terms of the states of their subsystems. For pure states {  of $N$-subsystems, described by a single ket $\ket{\Psi}$ acting in the composite Hilbert space $\mathcal{H}_1\otimes\dots\otimes\mathcal{H}_N$,} this fact manifests as correlations between some sets of local operators $O_1\otimes O_2\dots\otimes O_N\in\mathcal{B}(\mathcal{H}_1\otimes\dots\otimes\mathcal{H}_N)$ whose outputs cannot be reproduced by any local deterministic theory supplied with shared randomness. For pure states, the above description of quantum correlations encompasses both entanglement and nonlocality, { {the latter represented by the violation of a Bell inequality. It is well established that all entangled pure states violate a type of Bell inequality~\cite{Gisin1991,YuPRL2012}.}
It is known, however, that entanglement and nonlocality are not equivalent resources~\cite{Augusiak2015}, and there exist N-partite entangled states that do not violate a N-partite Bell inequality~\cite{ZukowskiPRL2002,ZukowskiPRA2002}.
} 
{ {For mixed states, describing mixtures of pure states $\ket{\Psi_i}$ and represented by a density matrix $\rho=\sum_i p_i \ket{\Psi_i}\bra{\Psi_i}$}}, classical and quantum correlations might compete, showing new facets of the non-classicality of the quantum state and allowing other forms of quantum correlations like quantum discord which  measures the amount of  ``non-classical" correlations embedded in the state~\cite{ZurekDiscord}. 

{ {A novel approach to understand the non-classical character of a quantum system is to consider it as a resource needed to perform a task which cannot be achieved if the state is classical.}} This concept allows one to understand in a unified frame different quantum correlations such as entanglement, steering, discord or nonlocality, together with other features arising from quantum superpositions, like quantum coherence, as quantum resources~\cite{DevetakIEEE2004}.

The theory of quantum correlations has also contributed to
the development of classical methods for simulating quantum many-body systems. A prominent example has been the insight {  into} the principles and applications of matrix product states (MPS) opening the route to the extension of the density matrix renormalisation group (DMRG) algorithm to time evolution, dissipative systems, mixed states and, using tensor networks, two-dimensional lattices~\cite{SchollwockMPS}.

Motivated by these developments, the goal of this report is to review the latest progress in the characterisation of strongly-correlated systems using quantum correlations. While recent reviews focus {  mostly on the analysis of bipartite entanglement in many body-systems}~\cite{LaflorencieReview}, which we briefly cover {  here} for completeness, we instead discuss more in depth other forms of quantum correlations like multipartite entanglement {  (entanglement shared between more than two subsystems)}, quantum nonlocality {  (impossibility of describing a composite system with a local hidden variable model)} and quantum discord {  (non-classical correlations arising also in some non entangled mixed states)}. The aim of this paper is not to review the complete theory behind these quantum correlations (for this we refer the reader to dedicated reviews~\cite{HorodeckiRMP2009,AditiReview,AdessoQuantumCorrelations,Braun2017}) but to showcase their applications for understanding condensed matter systems.

The plan for the review is as follows. In Sec.~\ref{sec:gallery}, we will make a short survey of the most common spin models that we will mention in the rest of the review. The reader familiar with interacting spin models can skip this section. In Sec.~\ref{sec:bipartite}, we will discuss the latest development on bipartite entanglement {  including the most employed measures to quantify entanglement in many-body systems, but also other tools like the entanglement spectrum and entanglement localisation. 
We will point out their significance in the analysis of quantum phase transitions. We will close this section by presenting  the so-called quantum marginal problem}. Then, we will turn the discussion to multipartite entanglement in Sec.~\ref{sec:multipartite}, introducing, first, two global {  entanglement} measures, namely geometric and global entanglement, and afterwards specialising onto genuine multipartite entanglement. Sec.~\ref{sec:nonlocality} will deal with the topic of quantum nonlocality in many-body systems. We will introduce the basic concepts {  regarding Bell inequalities and show their relevance in many-body systems}. Next, we will discuss the latest development on the detection of multiparticle quantum nonlocality using single and two-body correlators. We will also include some recent progress on temporal quantum nonlocality detected using the Leggett-Garg inequalities in out-of-equilibrium spin chains. In Sec.~\ref{sec:otherquantum}, we will discuss other quantum correlations that can be employed to characterise a quantum many-body state. These correlations include quantum discord, quantum  coherence-based correlations and measures based on the mutual information. At the end we will talk about the ultimate precision for estimating a parameter, for example a coupling constant or the temperature, of a many-body sample and the relation with quantum correlations. In Sec.~\ref{sec:conclussions} we {  will} summarise and conclude our report.  

\section{A small gallery of spin models}
\label{sec:gallery}

In this section we report the main features and phase diagrams of the models that are mostly discussed in the rest of this report. For more technical introductions to quantum many-body physics we refer the reader to specialised reviews or textbooks \cite{tsvelik2007quantum,schollwock2008quantum,auerbach2012interacting, Dutta2010}.

\subsection{Spin-1/2 models}
We start our gallery with a very general spin-1/2 Hamiltonian which  encompasses the most used models:
\begin{equation}
\label{eq:Hmodel}
H=J\sum_{\langle i,j\rangle}^{N} \left [\frac{1+\gamma}{2}\sigma_x^i \sigma_x^j+
\frac{1-\gamma}{2}\sigma_y^i \sigma_y^j +\Delta \sigma_z^i \sigma_z^j \right ] +B\sum_{i=1}^N \sigma_z^i
\end{equation}
 where the sum $\langle i,j\rangle$ is extended to nearest-neighbour spins depending on the topology of the underlying lattice and $N$ is the total number of spins.
 In Eq.~\eqref{eq:Hmodel}, $\sigma_{x,y,z}^i$ are the Pauli spin operators for site $i$; $J$ is the strength of the spin-spin coupling restricted to nearest-neighbours; $\gamma$ and $\Delta$ fix the anisotropy of the interactions in the $XY$ plane and along the $Z$ axis respectively. {  Finally, $B$ is an applied external} field along the $Z$ axis. 

\begin{figure}[t]
\begin{center}
\includegraphics[width=0.9\textwidth]{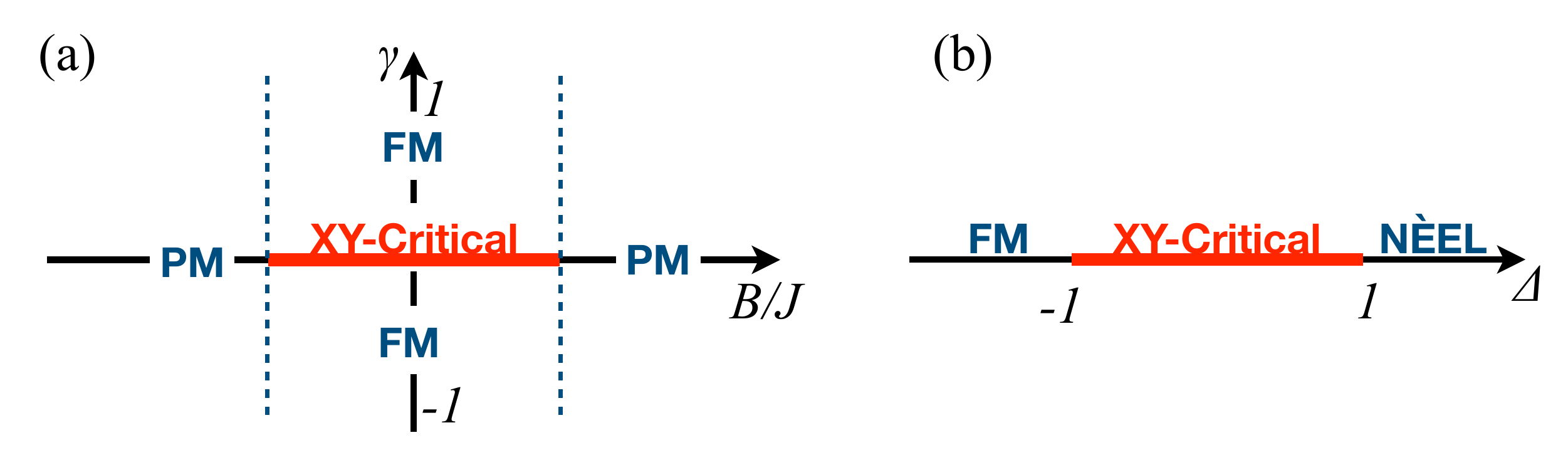}
\caption{(a) Phase diagram of the 1D XY model obtained from Eq.~\eqref{eq:Hmodel} for $\Delta=0$ and $J<0$. The thick horizontal red segment ($\gamma=0$, $|B/J|<1$) is the XY-critical phase; the system is in the ferromagnetic phase (FM) for $|B/J|<1$ and $\gamma\neq 0$ and in the paramagnetic phase (PM) elsewhere. The vertical dashed lines indicates second order phase transitions except for $\gamma=0$ for which a BKT transition separates the XY-critical and PM phases (see text)
(b) Phase diagram of the XXZ model obtained from Eq.~\eqref{eq:Hmodel} for $\gamma=B=0$. The system is in the FM phase for $\Delta<-1$, in the XY-critical phase for $|\Delta|<1$ and in the N\'eel phase for $\Delta >1$.
  }
\label{fig:phasediagrams}
\end{center}
\end{figure}

Despite the simple-looking form of Hamiltonian \eqref{eq:Hmodel}, the model cannot be solved exactly for generic lattices, especially for periodic lattices of dimensions $D>1$. In one-dimensional chains (1D), the ground state of the Hamiltonian \eqref{eq:Hmodel}, can be found exactly for various parameters. Let us distinguish two cases: 

(a) for $\Delta=0$, the model is generically known as the XY model which, in 1D,  can be diagonalised exactly { for both negative and positive values of $J$}  using the 
Jordan-Wigner transformation \cite{LSM,Pfeuty,Barouch1970,Barouch1971}. From this solution, the energy, correlation function and reduced density matrices of ground and thermal states can be efficiently obtained, including their time dependences in a real time evolution (for a review see \cite{PeschelEisler2009}). Therefore many measures of entanglement and quantum correlations can be calculated without difficulties. 
The phase diagram of the model is shown in Fig.~\ref{fig:phasediagrams}(a) for $J<0$. For $\gamma=0$, $|B/J|<1$, the ground state is gapless and belongs to the XY-critical phase characterised by quasi-long range order, i.e. spin-spin correlations functions decay algebraically. Otherwise the model is in the ferromagnetic phase (FM) for $|B/J|<1$ and $\gamma\neq 0$ and in the paramagnetic phase (PM) elsewhere. The lines $|B/J|=1$ are continuous phase transition between the FM and PM, except for $\gamma=0$, where the transition between the XY-critical phase and the PM phase is a multicritical factorisable point~\cite{BunderPRB1999}.
For $\gamma=1$, the Hamiltonian reduces to the Ising model in transverse field, which exhibits a second order phase transition between the ferromagnetic phase (FM) and the paramagnetic phase (PM).

(b) For $\gamma=B=0$ and $J>0$, the model is usually referred to as the XXZ model. This can be solved analytically using Bethe's ansatz, although quantum information quantities, e.g. entanglement and discord, are difficult to obtain in practice and one resorts to numerical simulations. As a function of the anisotropy $\Delta$, the phase diagram is shown in Fig.~\ref{fig:phasediagrams}(b). The system is in the FM phase for $\Delta<-1$, in the XY-critical phase for $|\Delta|<1$ and in the N\'eel phase for $\Delta >1$. The point $\Delta=-1$ coincides with a first order transition between the FM and XY-critical phases. At the point $\Delta=1$, also known as the isotropic Heisenberg point, a { Berezinskii-Kosterlitz-Thouless transition (BKT transition)} between the XY-critical and N\'eel phases occurs. 

Remarkably, the ground state of Hamiltonian  \eqref{eq:Hmodel} is completely factorised for the factorising field: $B_f=J\sqrt{(1+\Delta)^2-(\gamma/2)^2}$. For this particular value of the field, there are no correlations, quantum or classical, between spins \cite{Kurmann1982,RoscildePRL2005,GiampaoloPRL2010,Cerezo}. 

For generic values of the parameters, the fully anisotropic model in Eq.~\eqref{eq:Hmodel} is referred to as XYZ. In the rest, we will also discuss the case in which the strength of the interactions along the $X$ and $Z$ axis are equal, $\Delta=(1+\gamma)/2$, but different from the interactions along the $Y$ direction. This class of models is referred to as XYX.

For larger dimensional systems one can employ mean-field solutions to find the approximate ground state of model~\eqref{eq:Hmodel} that become more accurate as the dimensionality increases. In the limiting case of infinite dimension, corresponding to all spins interacting with each other, the model becomes exactly solvable. This is a particular case of the celebrated Lipkin-Meshkov-Glick (LMG) model from nuclear physics which can be rewritten as \cite{LMG}:
\begin{equation}
\label{eq:LMG}
H=\frac{J}{N} S_x^2 +B S_z
\end{equation}
 where $S_{x,z}=\sum_i \sigma_{x,z}^i$ are the total (collective) magnetisations along $X$ and $Z$ and the factor $N$ (number of particles)  is normally introduced to remove the divergence of the ground state energy per site when $N\to\infty$. The model for $J<0$ exhibits a second order quantum phase transition at $J=B$ of the mean-field universality class.

\subsection{Spin-1 models}
For spin-1 models the situation becomes richer than for the spin-1/2 case due to the competition of spin and quadrupolar orders and thanks to the emergence, in 1D, of symmetry-protected phases, like the Haldane phase. In this report we will discuss the quantum correlation properties of the Heisenberg Hamiltonian with uniaxial anisotropy of strength $U$
\begin{eqnarray}
\label{eq:bilin}
H = J\sum_{\langle i,j\rangle}^{N} (S_{x}^{i} S_{x}^{j} +S_{y}^{i} S_{y}^{j}  + S_{z}^{i} S_{z}^{j})+U \sum_{i}^N  \left(S_{z}^{i} \right)^2
\end{eqnarray} 
where $S_{x,y,z}^{i}$ are the spin-1 operators for site $i$.
The ground state phase diagram consists of three phases: for $U/J > 0.968$ the system is in the ``large D" phase in which the ground state is adiabatically connected to the $\ket{00\dots 0}$ state\footnote{In the literature the strength of the uniaxial field is usually denoted with $D$, hence the name ``large D". However to avoid confusion with discord, we will use $U$.}. For $U/J< -0.315$ the system is in the N\'eel antiferromagnetic phase, characterised by the staggered magnetisation. For the intermediate values of $U/J\!\in\![-0.315,0.968]$ the system is in the Haldane phase. The entire phase is gapped, characterised by free edge spins and a non vanishing string order parameter \cite{Rommelse, Schollwock1996}:
\begin{equation}
\label{eq:string}
O = \lim_{r\to\infty} \left\langle S_{z}^i\exp\left[i\pi\sum_{j=i+1}^{i+r-1} S_{z}^j \right]S_z^{i+r}\right\rangle.
\end{equation}
The Haldane phase can also be characterised by the structure of the bipartite entanglement spectrum which will be discussed in the next section.

%% file: Chapter2.tex

\section{Bipartite entanglement in many-body systems}
\label{sec:bipartite}

The crucial role of bipartite entanglement in many-body systems is, at least, four-fold. 
First, the fact that many-body Hamiltonians are local, i.e., $H=\sum H_k$, {  where each term, $H_k$,} has at most $k$-body interactions, forces bipartite entanglement of gapped ground states in 1D to follow an area law. Generic multipartite pure states follow instead a volume law. Second, bipartite entanglement is the pivotal concept to develop the building blocks of tensor network structures, e.g. matrix product states, corresponding to accurate representations of ground states for a large class of  many-body quantum correlated systems. Third, since physical interactions in Nature are usually two-body, bipartite entanglement provides an alternative and novel characterisation of quantum phase transitions. Finally, topological quantum matter is characterised by long range entanglement that can be identified with subleading terms in the bipartite entanglement, the so-called topological entanglement.

Recently, valuable reviews about bipartite entanglement in many-body systems have appeared in the literature. Some of them put the emphasis on quantum information \cite{LewensteinOxford2012,SchuchPRB2011,Zeng2016}, others have a condensed matter physics flavour \cite{LaflorencieReview,AmicoRMP2008}, and some others take a computer science perspective \cite{AharanovIEEE2011}. We refer the reader to previous reviews for a detailed insight on the subject of bipartite entanglement in quantum matter. For that reason, here we restrict ourselves to a brief description of the most used bipartite entanglement concepts in many-body systems. 

When dealing with many-body strongly correlated systems, there are at least two different types of questions in which bipartite entanglement plays a crucial role. The first type aims at finding the ground state ($\ket{\Psi_0}$) or low energy sector properties of a certain Hamiltonian $H$ as given by the  expectation values of some observables $\bra{\Psi_0} O_i\ket{\Psi_0}$. For a large class of Hamiltonians, the problem of finding the ground state itself is known to be quantum NP-hard, more precisely it is a problem in the complexity class called Quantum Merlin Arthur (QMA)~\cite{SchuchPRL2008,AharanovCMP2009}, so not even a quantum computer could efficiently solve it. Accurate approximations of ground states are provided by tensor networks like matrix product states (MPS), projected entangled product states (PEPS) or multi-scale entanglement renormalisation ansatz (MERA). Alternatively, one can construct ad-hoc an $N$-partite quantum state $\ket{\phi}$ with desired properties (entanglement scaling law, symmetries, topological properties, \dots) and search for its parent Hamiltonian $H_p$, so that  $H_p\ket{\phi}=E_0\ket{\phi}$. This question is related also to the quantum marginal problem and the $N$-representability problem of many-body systems, which we will discuss later in this section.
The second type of problems which involves entanglement is the classification and nature of the different quantum phases present in strongly correlated systems. {  Quantum phase transitions which fall under the Landau-Ginzburg paradigm of phase transitions, are associated to the spontaneous breaking of a symmetry and classified  according to the minimum order of the derivative of the ground state energy which is not continuous. However, the irruption of topological and other exotic quantum phase transitions which do not break any local symmetry, demands for a  broader definition of  quantum phase transitions}. We adopt the general definition---which does not rely on symmetry arguments---that two local gapped Hamiltonians $H_1$ and $H_2$ describe states in the same quantum phase if they can be connected  by a smooth path of Hamiltonians which keeps the Hamiltonians local and the gap open \cite{XiaoGangWen2017}. Different quantum phases will correspond to Hamiltonians which cannot be connected by such a smooth path. Accordingly, quantum phase transitions correspond to abrupt changes of the ground state properties and are signalled by the non analytical behaviour of the ground state energy or other observables.

\subsection{Bipartite entanglement measures}
\begin{figure}[t]
\begin{center}
\includegraphics[width=0.5\columnwidth]{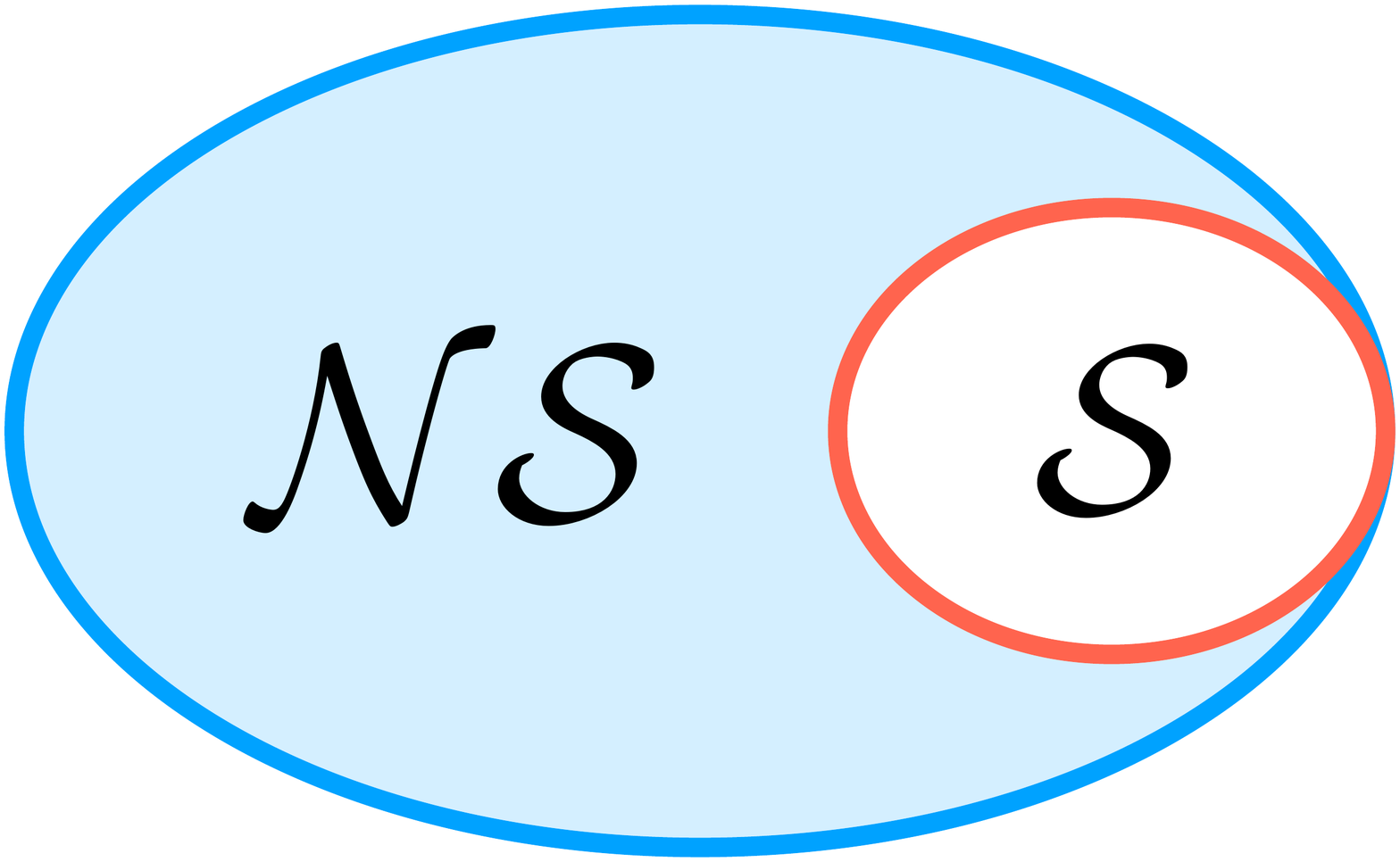}
\caption{Schematic drawing of the convex set of separable states $\mathcal S$ and the (not convex) set of entangled or nonseparable states $\mathcal{NS}$.}
\label{fig:setentangled}
\end{center}
\end{figure}

Entanglement is a quantum correlation which can be easily characterised. A composite quantum system of $N$ subsystems {  described by a density matrix}  $\rho_{1,2,\dots N}\in \mathcal{H}_1\otimes\mathcal{H}_2\dots\otimes\mathcal{H}_N$, {  where $\mathcal{H}_i$ denotes the Hilbert space of party $i$}, is entangled iff it cannot be written as a convex combination of the form 
\begin{equation}
\rho_{1,2... N}=\sum_i p_i\, ( \rho_i^1\otimes\rho_i^2...\otimes\rho_i^N )
\label{separable}
\end{equation} 
with $\sum_i p_i=1$ and $\forall p_i>0$. In general, since the above decomposition is by no means unique, to determine if a generic quantum state is separable or entangled is a difficult task, unless the state is pure, i.e., $\rho_{1,2\dots N}=(\ket{\Psi}\bra{\Psi})_{1,2\dots N}$. {  States that can be expressed as (\ref{separable}) are called separable. They form a convex set denoted usually by $\mathcal S$ (see Fig.~\ref{fig:setentangled}). Pure separable states are also called product states.}

Entanglement measures (bipartite, multipartite) are single quantities that serve to certify the presence/absence of entanglement but also to quantify entanglement as an operational resource to perform tasks that are impossible if the quantum state is separable~\cite{HorodeckiRMP2009}. An entanglement measure, $\mathcal{E}$, must verify the very logical assumption that it cannot increase under operations made locally on the different parts of the composite system even if classical communication is shared between the different subsystems. In other words, any measure of entanglement should be  monotonically decreasing under maps $\Lambda$, {  arising from composition of local operations and classical communications (LOCC)} i.e., $\mathcal{E}(\Lambda(\rho))\leq \mathcal{E}(\rho)$. There are other properties like additivity, strong monotonicity,$\dots$ that are desirable properties but not mandatory for an entanglement measure to be meaningful. The condition that separable states have zero entanglement easily derives from the monotonicity condition. 

Except for very few instances, there is not a unique measure of bipartite entanglement. However, for  pure states and only for those, any bipartite splitting $A/B$ admits---independently of the size of the {  partition and 
the physical dimension of the subsystems}---the so-called Schmidt decomposition which links the global pure state with the spectrum of its reduced density matrices { (marginals)}. 
Formally, every $\ket{\psi}_{AB}$ acting on ${\cal H}_{A} \otimes {\cal H}_{B}$ can be represented in an appropriately chosen basis as:
\begin{equation}
\ket{\psi}_{AB} = \sum _{i=1}^{M} \sqrt{\lambda_i} \;(|e_i\rangle \otimes |f_i\rangle),
\label{eq:schmidt}
\end{equation}
where $\{\ket{e_i}\}$($\{\ket{f_i}\}$) forms an orthonormal basis in
${\cal{H}}_{A}$(${\cal{H}}_{B}$), $\lambda_i> 0$, and  $\sum_{i=1}^{M} \lambda_i=1$, where $M\le \dim{\cal H}_{A}, \dim{\cal H}_{B}$.
The positive real numbers, \(\lambda_i\), are known as the Schmidt coefficients, (also Schmidt spectrum, Schmidt eigenvalues, entanglement spectrum) and the vectors $\{\ket{e_i} \otimes \ket{f_i}\}$ are the Schmidt eigenvectors of \(|\psi_{AB}\rangle\). Note that uncorrelated (product) pure states correspond to those whose Schmidt decomposition has one and only one Schmidt coefficient. If the decomposition has more than one Schmidt coefficients, the pure state is entangled. 
Importantly, the reduced density matrices with respect to the partition $A/B$ are given by  $\rho_A=\sum_i\lambda_i\ket{e_i}\bra{e_i}$ ($\rho_B=\sum_i\lambda_i \ket{f_i}\bra{f_i})$, so the Schmidt coefficients are the eigenvalues of both reduced density matrices, which are basis independent. Notice that for a pure state, like the ground state of a many-body system, the information contained in the Schmidt decomposition is, by default, complete with respect to the corresponding bipartite splitting.\\

\noindent {\it{Entanglement entropy.}} There are two measures of bipartite entanglement conceptually very important which lead to 
the definition of entanglement entropy. The latter is the unique measure of bipartite entanglement for pure states which is operationally meaningful. The first one, is the entanglement of distillation, $\mathcal{E}_D$, of a given bipartite state $\rho_{AB}$,  which is defined as the optimal rate of maximally entangled states (singlets for spin-1/2 systems) that can be distilled from it using LOCC~\cite{BennettPRA1996,BennettPRL1996}. The measure is defined in the asymptotic limit, so having at disposal an infinite number of copies of the state $\rho_{AB}$. The complementary measure is the entanglement cost, $\mathcal{E}_C$,  defined as the rate of maximally entangled states (singlets) needed to create $\rho_{AB}$ in the asymptotic limit using LOCC~\cite{HaydenJPA2001}. Any other measure of entanglement, $\mathcal{E}$, is lower bounded by the entanglement of distillation and upper bounded by the entanglement cost~\cite{Horodecki2000,Donald2002}: $\mathcal{E}_D\leq \mathcal{E} \leq \mathcal{E}_C$. 
These measures, despite their operational meaning, have little applicability in general because they involve optimisation over the LOCC set, which is a hard computational problem. However, for pure states and only for them, the entanglement cost and the entanglement of distillation coincide and are given by the von Neumann entropy of the subsystems~\cite{BennettPRA1996}:
\begin{equation} 
\mathcal{E}_E(\ket{\psi_{AB}}) = S(\rho_A) =S(\rho_B),
\label{ent_entropy}
\end{equation}
where the von Neumann entropy of a state is given by  $S(\rho)=-\Tr\rho\log(\rho)$. Making use of the Schmidt decomposition, the von Neumann entropy can be expressed as $\mathcal{E}_E(\ket{\psi_{AB}})=-\sum_i\lambda_i\log(\lambda_i)$. The quantity $\mathcal{E}_E(\ket{\psi_{AB}})$ is often called the ``entropy of entanglement", or simply, the entanglement of $\ket{\psi_{AB}}$. This fact holds for all pure states, independently of the dimension of the corresponding Hilbert space ${\cal{H}}_{A}$(${\cal{H}}_{B}$) {  after the partition, and the dimensionality of each physical subsystem. For example, given a system of $N$ spins-1/2 particles, there are many possible bipartitions of the full system: $k$ spins versus $N-k$ spins for $1\le k< N$. For each partition it is possible to express the pure state with a Schmidt decomposition, where the dimension of ${\cal{H}}_{A}$ (${\cal{H}}_{B})$ depends on the chosen partition and so they do the corresponding reduced density matrices and thus the
entanglement entropy.}\\

\noindent {\it{Renyi Entropies.}} Another bipartite entanglement measure of the state $\ket{\psi_{AB}}$ often used in many-body systems are Renyi entropies, $S_n(\rho_A)$,  defined as:
\begin{equation}
S_n(\rho_A) =\frac{1}{1-n}\log\Tr{\rho_A^n},
\end{equation}
which are entanglement measures for mixed states for any real number $n\ge 0$ and $n\neq 1$. Like the entanglement entropy, these measures depend on the eigenvalues of the reduced density matrices, although in a stronger non linear way. It is interesting to notice that while for the entanglement entropy, $S(\rho_A)$, the Schmidt eigenvalues which are in the middle of the entanglement spectrum are the ones with larger weight, for the Renyi entropies, $S_n(\rho_A)$, different sectors of the entanglement spectrum are enhanced depending on the real number $n$. \\

\noindent\emph{Bipartite entanglement measures for mixed states.}
For pure states, all measures of bipartite entanglement are in one-to-one correspondence and all are a function of the eigenvalues of the reduced density matrix arising from the chosen partition \cite{HorodeckiRMP2009,VidalPRA2002}. For mixed states, this is not anymore the case, with the notable exception of mixed states of two spin-1/2 particles or qubits, where there exists a measure called concurrence, $\mathcal{C}$, which is equivalent to the entanglement cost and is defined as~\cite{WoottersPRL1998}
\begin{equation}
\mathcal C=\max(0,\mu_1-\mu_2-\mu_3-\mu_4)
\label{eq:concurrence}
\end{equation}
where $\mu_i$ are  the eigenvalues of the matrix $R=\sqrt{\rho\rho'}$ with $ \rho'=\sigma^y\otimes \sigma^y\rho^*\sigma^y\otimes\sigma^y$.

There are other pairwise entanglement measures that are used to calculate entanglement also beyond spin-1/2. One of them is the so called negativity, $\mathcal{N}$, which is a very simple measure based on the partial transpose of the density matrix with respect to one of the subsystems and it is defined as $\mathcal{N}=\sum_i |\tilde{\mu}_i|$,
where now $\tilde{\mu}_i$ refers to the negative eigenvalues of the partial transpose matrix $\rho^{T_A}$~\cite{VidalPRA2002,ZyczkowskiPRA1998}. For a given state $\rho_{AB}\in\mathcal{H}_A\otimes \mathcal{H}_B$, {  with Hilbert space dimensions $m$ and $n$, respectively}, the partial transpose matrix is defined as $\rho^{T_A}:=(T\otimes I)\rho_{AB}$, where $T$ is the transpose action acting on subsystem $A$, and, $I$, the identity operator acting on B. The maximal number of negative eigenvalues of the partial transpose matrices can never be larger than $(m-1)(n-1)$ ~\cite{Rana2013}. Negativity has recently been used for quantifying entanglement of thermal states of lattice fermions in 1D~\cite{ParkPRL2017}. Other measures, relying to geometrical distances between different quantum states are also powerful, both in the bipartite as well as in the multipartite case. The most known is the relative entropy of entanglement \cite{VedralPRL1997} defined as {  the closeness of a given state $\rho$ to a state in $\mathcal S$:} 
\begin{equation}
\label{eq:relative_entropy_entanglement}
\mathcal{E}_R=\min_ {\sigma\in \mathcal{S}} S(\sigma||\rho),
\end{equation}
where 
\begin{equation}
\label{eq:relative_entropy}
S(\sigma||\rho):=\Tr(\rho\log\rho-\rho\log\sigma)
\end{equation}
is the quantum relative entropy or quantum  Kullback-Leibler  divergence. The quantum relative entropy is not a distance because is not symmetric and does not obey the triangle inequality. When the set $\mathcal{S}$ corresponds to the set of separable states, the relative entropy of entanglement leads to the geometrical entanglement, that we will discuss in detail in Sec.~\ref{sec:multipartite}. Finally, it is worth mentioning the robustness of entanglement introduced by Vidal and Tarrach~\cite{VidalPRA1999} which is defined as the minimal amount, {  $s$, of a separable state $\sigma$  (acting as noise)}  that can be added to an entangled state $\rho$ before it becomes a separable state:
\begin{equation}
\label{eq:robustness}
R(\rho)=\min_ {\sigma\in \mathcal{S},s\in\mathbb{R}} \left\{s:\frac{\rho+s\sigma}{1+s} \in\mathcal{S} \right\}.
\end{equation}
We will discuss its applicability to many body systems in Sec \ref{sec:multipartite}.

In special cases it is possible to compute entanglement in many-body systems at non-zero temperature. In Ref.~\cite{LeePRL2015}, a single and two-channel Kondo model is considered and the entanglement of formation of an impurity with the rest of the system is computed using a minimisation over entanglement witnesses.\\

\noindent\emph{Entanglement localisation.} In contrast to the bipartite entanglement measures presented above, the entanglement localisation does not focus on the pairwise entanglement between two nearby sites as it is normally done, but introduces the idea of ``long range" entanglement. Typically, for the spin chain models of Sec.~\ref{sec:gallery}, the concurrence \eqref{eq:concurrence} between spins $i,j$ far away ($|x_i-x_j|\gg 1$) is simply zero, indicating that there is not pairwise entanglement between them. For a given state $\rho_{1,2,\dots N}$, the localisable entanglement $\mathcal{E}_{i,j}$ is defined as the maximum entanglement that can be created (localised) on average between spins $i$ and $j$ by performing local measurements (and classical communication) on the other spins~\cite{VerstraetePRL2004a,VerstraetePRL2004,PoppPRA2005}. The entanglement localisation leads to the concept of entanglement localisation length, $\xi_E$,  which is the typical length scale at which it is possible to create maximally entangled states by performing local measurements in the reminding parties. The entanglement localisation length has been used to detect topological phases which cannot be detected with other bipartite entanglement measures~\cite{VerstraetePRL2004}.

\subsection{Local Hamiltonians and the area law} 

A very important consequence of the locality of many-body Hamiltonians is the scaling of the bipartite entanglement of a region versus its complementary, as measured by the entanglement entropy $S(\rho_A)$~\cite{EisertRMP2010,HastingsLesHouches2010}. For ground states of local Hamiltonians in $D$-dimensions, the entanglement entropy between a region of  size $L$ and its complement goes like the boundary between them, $S\sim L^{D-1}$ (and not like the volume $S\sim L^{D}$). Corrections to this area law appear in critical points (gapless), where the entanglement entropy increases logarithmically with the size of region~\cite{HolzheyNPB1994,KorepinPRL2004,CalabreseJSTAT2004,VidalPRL2003}, and in topological systems, where there is an additive correction to the area law, independent of the size of the region, indicating long-range entanglement~\cite{KitaevPRL2006}. The area law has been rigorously proven for gapped 1D systems \cite{HastingsJSTAT2007}, and for thermal states regardless the dimension of the system \cite{WolfPRL2008}.  The existence of such an area law implies that ground states of all 1D local Hamiltonians can be approximated with the desired accuracy with an MPS~\cite{HastingsLesHouches2010}. Therefore, the ground state of every gapped spin chain model is well approximate by an MPS. The accuracy of the approximation depends, obviously, on the bond dimension of the MPS, {  that quantifies the entanglement of the auxiliary systems in which the MPS is built on}. Finding the energy of a Hamiltonian whose MPS bond dimension grows polynomially with the size of the system is an NP-hard problem~\cite{SchuchPRL2008}. A detailed review of the derivations of the area law for higher dimensions, topological states and fermionic systems is provided in \cite{LaflorencieReview}.

\subsection{Entanglement spectrum}
\begin{figure}[t]
\begin{center}
\includegraphics[width=\columnwidth]{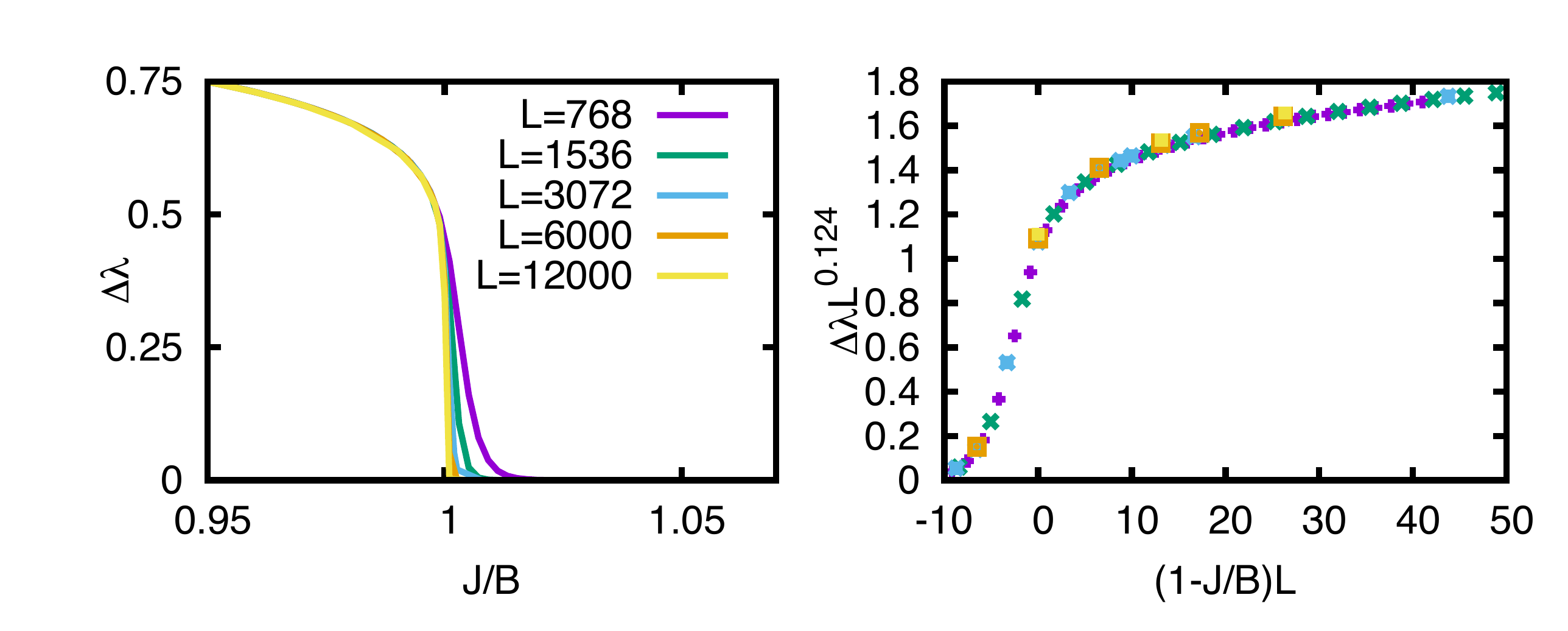}
\caption{Left: Schmidt gap $\Delta\lambda$ for the ground state of the Ising chain with open boundary conditions as a function of $J/B$ for chain lengths: $L=768,1536,3072,6000,12000$. Right: scaling and collapse of the Schmidt gap with fitted critical exponents: $\beta=0.124$ and $\nu=1.0$. Adapted from Ref.~\cite{DeChiaraPRL2012}.}
\label{fig:schmidt}
\end{center}
\end{figure}

Beyond the entanglement entropy or any other entanglement measure, further characterisation of many-body systems can be obtained if the whole entanglement spectrum resulting from a bipartition of the system is investigated. Li and Haldane~\cite{LiPRL2008} and, immediately later, various authors \cite{PollmannPRB2010} pointed out that the full entanglement spectrum, $\{\lambda_i\}$, (Schmidt coefficients) arising after a cut, shows features linked to ground properties that are blind to any entanglement measure (being a single number). Thus, the distribution of the Schmidt eigenvalues is related to the Hamiltonian symmetries, to edge states and to quantum phase transitions. In particular, in 1D these features are relevant to detect topological phases. For instance, Pollmann et al. \cite{PollmannPRB2010}  showed that in spin-1 chains, some topological phases (e.g. Haldane phases) are characterised by an even degeneracy of the full entanglement spectrum and this characterisation applies equally for
topological phases that have neither gapless edge modes nor string
order. Further, in 1D systems, it was established that the degeneracies shown in the entanglement spectrum are linked to the symmetries that  protect the different quantum phases. This relation extends even further at  phase transitions where a direct link associates the entanglement spectrum to the conformal field theory describing the former and allows to classify all 1D gapped phases of local Hamiltonians~\cite{CalabreseCardyReview2009,SchuchPRB2011}. For a gapped topologically ordered system, the entanglement spectrum reproduces the low excitation of the edge states.
The entanglement spectrum shows also an intricate structure at quantum phase transitions, where several Schmidt eigenvalues cross.  Also, 
the entanglement gap between the largest non trivially degenerated Schmidt eigenvalues, the so-called Schmidt gap, has been proposed as a nonlocal parameter which acts as a precursor of quantum phase transitions in quantum spin chain models including topological phases \cite{DeChiaraPRL2012,LeporiPRB2013}, and applied also in the quantum impurity model~\cite{BayatPRL2017} and in the many-body localisation problems~\cite{Gray2017}. { An example of the scaling of the Schmidt gap, denoted by $\Delta\lambda$ is shown in Fig.~\ref{fig:schmidt} for the ground state of the Ising Hamiltonian obtained from Eq.~\eqref{eq:Hmodel} for $\gamma=\Delta=0$. It is observed that for $J/B<1$, in the paramagnetic phase the ground state is non degenerate and the Schmidt gap decreases to zero as $J/B$ approaches the quantum phase transition to the ferromagnetic phase extending for $J/B>1$. For this model, numerical evidence shows that the Schmidt gap follows a universal scaling of the form: $\Delta\lambda = L^{-\beta/\nu}f[(1-J/B)L^{1/\nu}]$ where $\beta=1/8$ and $\nu=1$ are the critical exponents of the Ising chain. The Schmidt gap can be also employed as a critical quantity in time-dependent problems, for example for sudden quenches and slow driving through equilibrium quantum phase transitions~\cite{CanoviPRB2014,TorlaiJSTAT2014,HuPRB2015}.} In larger dimensions, the universality properties of the entanglement spectrum is, however, a highly debated question \cite{ChandranPRL2014,Moreno-CardonerJSTAT2014}. 

\subsection{Bipartite entanglement and quantum phase transitions}

 First studies of the entanglement behaviour in simple spin models like the quantum Ising spin-1/2 chain {  (see Eq.~\eqref{eq:Hmodel})}, were performed by analysing pairwise entanglement between two adjacent spins, as measured by the concurrence as a function of the external magnetic field~\cite{Osterloh02,Osborne02}. Near the second order quantum phase transition occurring in this model, the concurrence is a continuous function but its first derivative is singular at the critical point in the thermodynamical limit. Using finite size scaling (FSS), the concurrence allows one to obtain the correct critical exponents corresponding to the Ising transition. These seminal contributions opened a new path in the analysis of QPTs using entanglement. Since then, a large amount of work has been devoted to deepen the connections between bipartite entanglement and QPTs, see for instance~\cite{AmicoRMP2008,VidalPRL2003,LatorreQIC2004,Alcaraz04,WuPRL2004}.

The connections between bipartite pairwise entanglement and quantum phase transitions can be brought further by realising that the ground state energy of any local Hamiltonian depending on a set of parameters $\alpha$,  $H(\alpha)=\sum H_k(\alpha)$, can be expressed through minimisation over its $k$-order reduced density matrices $\rho^{k}$ {  (marginals) } acting on the support of the local Hamiltonian $H_k$: 
\begin{equation}
\label{Kohn}
E_0(\alpha)=\min_{\ket{\Psi_0}}\bra{\Psi_0}H(\alpha)\ket{\Psi_0}=
\min_{\rho^k,\ket{\Psi_0}}\sum_{k} {\rm Tr}\;\left({H}_{k}(\alpha)\rho^{k}\right).
\end{equation}
Typically, Hamiltonians involve just two-body interactions  ${H}_k={H}_{ij}$. In this case, it follows trivially that $\partial_{\alpha} E_0\sim(\partial_\alpha H_{ij}) \rho^{ij}$. If the local Hamiltonians are smooth functions of the parameters, then a one-to-one correspondence can be made between the discontinuities of the ground state energy and the behaviour of the matrix elements of $\rho^{ij}$. Thus, pairwise entanglement measures which depend exclusively on $\rho^{ij}$ should be discontinuous in a first order quantum phase transition. By the same reasoning, it can be shown that a {  a second order quantum phase transition} should display a singularity in the derivative of these pairwise entanglement measures. Wu, Sarandy and Lidar \cite{WuPRL2004} summarised the above results stating that, for quantum phase transitions falling under the paradigm of symmetry breaking, the order and properties of quantum phase transitions of local Hamiltonians are signalled by entanglement measures depending on the corresponding reduced density matrix of the ground state. The theorem works in both directions, i.e., a discontinuity in a pairwise measure of entanglement in a 2-local Hamiltonian indicates a first order quantum phase transitions while a discontinuity/divergence in its derivative signals a second order one. For ground states obtained numerically in finite lattices, quantum phase transitions near multicritical points and some models~\cite{YangPRA2005,GuPRA2013} the above correspondence can fail. We remark again that only quantum phase transitions associated to discontinuities of the ground state energy allow to establish such a correspondence.

\subsection{Bipartite entanglement and topological order}
Topological order is a fascinating state of matter whose definition remains still controversial. Topological order is associated to highly degenerated disordered gapped ground states of quantum matter which do not break any symmetry and are robust agains local perturbations. A way to characterise such elusive states is by analysing their departure from the area law  $S(\rho)^{D-1}- \gamma$. This additive correction, $\gamma$, is normally referred to as the topological entanglement~\cite{KitaevPRL2006,LevinPRL2006} and is independent of the size of the blocks of partition, showing the long range nature of the topological entanglement. As it is often the case (with the exception of topological mean field models like Kitaev's model) many topological states correspond to the ground state of a disorder and frustrated Hamiltonian in dimensions $D>1$. In such cases, the ground state has to be found numerically, and to detect explicitly the topological correction to the entanglement entropy is a hard task. For these reasons, alternative methods to estimate the scaling behaviour of the entanglement entropy of a ground state which is a superposition of dimer coverings have appeared in the literature. Among them, the valence bond entanglement entropy \cite{AletPRL2007,ChhajlanyPRL2007}  which is defined as the average number of singlets that are crossed along one partition of the system, has been used to predict scaling properties in the different antiferromagnetic phases of the 2D Heisenberg spin-1/2 model. An alternative insight on the topological order in 1D chain and ladders can also be obtained using multipartite entanglement measures.

\subsection{The quantum marginal problem and quantum de Finetti's theorem} 
We finish this section by revisiting the quantum marginal problem, 
and the, closely related, quantum de Finetti's theorem. In the previous point, we have seen that the ground state energy of a many-body local Hamiltonians can be calculated via reduced density matrices which act on the local support of the corresponding $H_k$. Since the number of parameters of $\rho^k$ are, in principle, much smaller than the number of parameters of the $N$-body ground state itself $\ket{\Psi_0}$, one might wonder if the problem of finding the ground state energy can be solved the other way round. In other words, one can ask if it is sufficient to minimise over the reduced density matrices to determine the ground state energy. The answer is that minimisation over the reduced density set is as hard (in general NP-Hard) as finding the ground state of a local Hamiltonian, because minimising over the set of reduced density matrices (marginals) of a given ground state is not equivalent to minimising over the set of all density matrices of the same dimension. Unless there are some symmetries, minimisation over marginals remains hard. The quantum de Finetti's theorem indicates when the ground state of a $N$ many-body interacting Hamiltonian can be approximated by a product state. This is the very spirit of the Hartree-Fock approximation, where an interacting many-body Hamiltonian admits a mean field solution in which all particles are in the same state. A possible way to approach quantum de Finetti's theorem goes as follows: which are the marginals e.g. $\rho_{ij}$ which admit an $s$-copy symmetric extension in the limit $s\rightarrow \infty$ ? {  A symmetric extension means finding an s-party pure state, $\ket{\Psi}_s$, that is symmetric under permutation of the
extended parties and such that its marginals are exactly $\rho_{ij}$, i.e.
$\tr_k (\ket{\Psi}_s\bra{\Psi})=\rho_{ij}$ where $k$ comprises all subsystems except $i,j$.}  Of course, if the state admits an $s$-copy extension, it also admits an $s-1$ copy extension. It happens that there is only one set in which all symmetric extensions are allowed: the set of separable states. The quantum de Finetti's theorem then states that the $k$-reduced density matrices of an $N$-particle permutationally invariant state (bosonic) can be approximated---with an error that goes at most $O(k m^2/N)$---by a mixture of product states $\ket{\alpha}^{\otimes k}$, where $\ket{\alpha}$ is a single particle wavefunction and $m$ is related to the dimension of the particles \cite{Christandl2007}. So, in the thermodynamic limit, $N\rightarrow\infty$, the error goes to zero.
Notice that the theorem makes no assumptions on the Hamiltonian but on the permutational symmetry of the ground state. From here and using Eq.~\eqref{Kohn}, it follows trivially that the ground state energy of a generic bosonic interacting system given by the Hartree-Fock approximation is exact in the thermodynamic limit, even if the ground state itself is not a product state~\cite{Zeng2016}.

%% file: Chapter3.tex

\section{Multipartite entanglement}
\label{sec:multipartite}


\begin{figure}[t]
\begin{center}
\includegraphics[width=0.7\columnwidth]{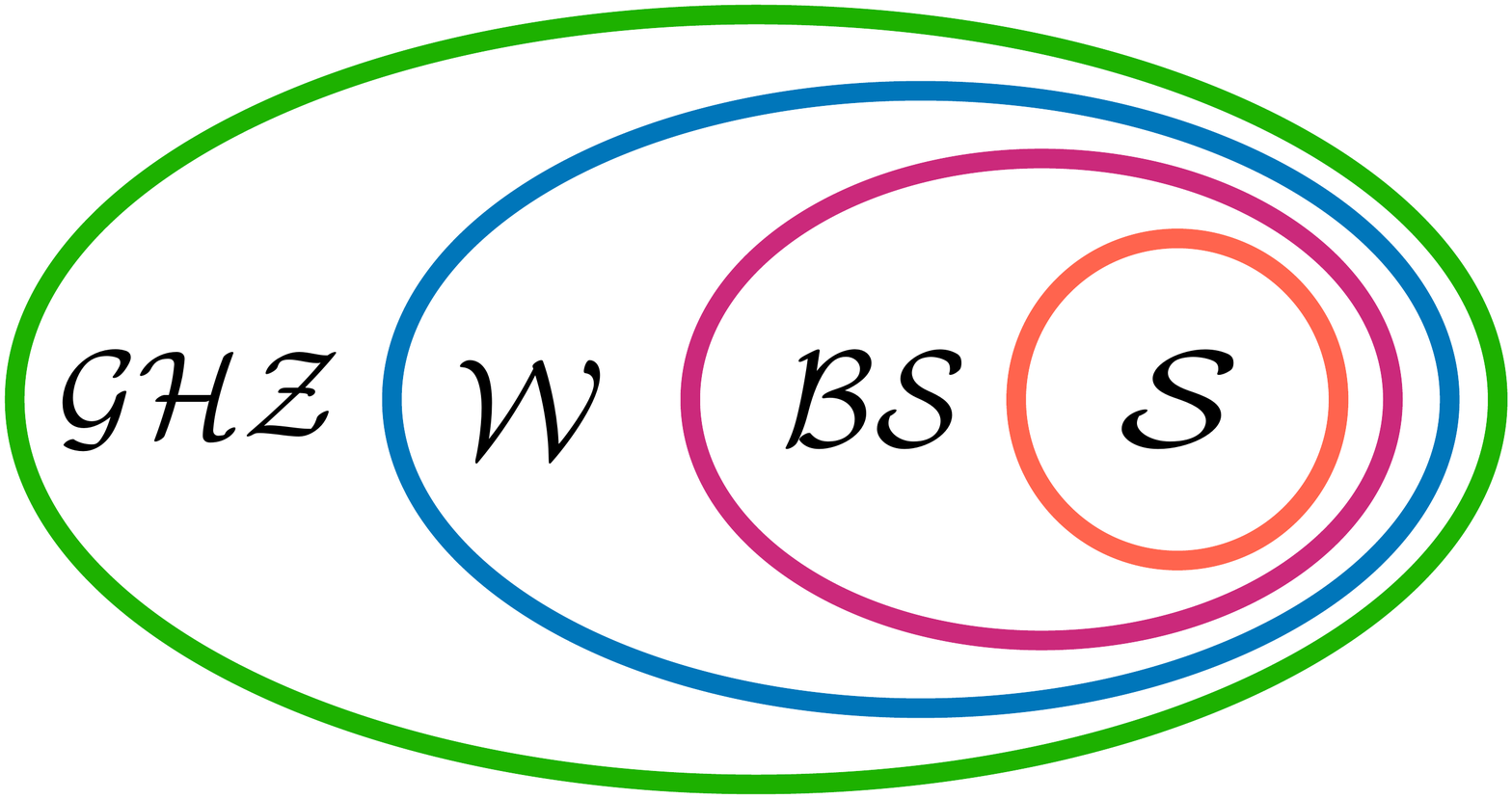}
\caption{Structure of entangled states classes of three qubits: separable states $\mathcal S$, biseparable states $\mathcal BS$, $\mathcal{W}$ states and $\mathcal{GHZ}$ states.}
\label{fig:multipartite}
\end{center}
\end{figure}
In the previous section, we have discussed bipartite entanglement in many-body systems for which entanglement is defined between two subsystems. Here the subsystem can refer to individual spins, to sets of spins or, in general, to sets of degrees of freedom in which the Hilbert space is partitioned. In the case of more than two parties, entanglement becomes more complicated to characterise, as it is possible for more than two parties to be entangled in different inequivalent ways. { For example, there are four inequivalent classes for a generic state of three qubits
~\cite{Dur2000,Acin2001}:
(i) separable states $\mathcal S$  of the form $\rho=\sum_i p_i\,
\rho_{i}^{(1)}\otimes \rho_{i}^{(2)}\otimes\rho_{i}^{(3)}$, where $p_i$ are probabilities and $\rho_{i}^{(1,2,3)}$ are density operators for qubits 1,2,3, respectively; (ii) biseparable states
$\mathcal B$ that can be obtained as linear combinations of states separable with respect to one of the
partitions $1|23$, $2|13$ or $3|12$; two so-called genuine multipartite entangled classes (iii) $\mathcal{W}$ states, and (iv) $\mathcal{GHZ}$ states. 
Each class contains those that are lower in the hierarchy, i.e. $\mathcal S\subset \mathcal B
\subset \rm{W} \subset \rm{GHZ}$ as shown in Fig.~\ref{fig:multipartite}. The distinction between the $\mathcal{W}$ states and the
$\mathcal{GHZ}$ ones arises from the fact that for, three qubits, there are two non-equivalent
classes of genuinely entangled states with representative elements being precisely the
W state $ \ket{\rm W}=1/\sqrt{3}(\ket{100}+\ket{010}+\ket{001})$, and the GHZ state
$\ket{\rm GHZ}=1/\sqrt{2}(\ket{000}+\ket{111})$. Elements of one class cannot be
inter-converted into elements of the second one using stochastic local operations and
classical communications. Therefore, the $\mathcal{W}$ and $\mathcal{GHZ}$ classes are formed by convex
combinations of states equivalent to $\ket{\rm W}$ and of combinations of states
equivalent to $\ket{\rm GHZ}$, respectively.
}

For detailed introductions to multipartite entanglement we refer the reader to specialised reviews \cite{HorodeckiRMP2009,GuhneTothPhysRep2009}. 
In this section we will discuss different measures of multipartite entanglement applied to strongly correlated systems. We start with geometric entanglement and global entanglement, which quantify how much a system differs from a fully separable state.  In Sec.~\ref{sec:genuine}, we discuss the presence of genuine multipartite entanglement which is a collective form of entanglement which cannot be reduced to bipartite entanglement only.

\subsection{Geometric entanglement}

Geometric entanglement (GE) is a collective measure of entanglement for multipartite states originally defined in Refs.~\cite{Shimony1995,BarnumLindenJPA2001} and later extended to mixed states and applied in the context of quantum many-body states in Refs.~\cite{WeiGoldbartPRA2003, WeiPRA2005}.
It represents the  geometric distance of a given multipartite state from its closest product state. Following Ref.~\cite{WeiGoldbartPRA2003}, let us consider a generic pure quantum state of $N$ parties:
\begin{equation}
\ket\psi = \sum_{i_1,i_2,\dots,i_N} \Psi_{i_1,i_2,\dots,i_N} \ket{i_1,i_2,\dots,i_N}
\end{equation}
where $\{\Psi_{i_1,i_2,\dots,i_N}\}$ are the state expansion coefficients in an arbitrary basis of orthonormal  product states $ \ket{i_1,i_2,\dots,i_N}$. 

Now, let us consider the overlap between the state $\ket\psi$ and the product states $\ket{\phi_{\rm prod}}=\otimes_{i=1}^N \ket{\phi_i}$ and maximise over the set of product states,
\begin{equation}
\Lambda_{\rm max} = \max_{\ket{\phi_{\rm prod}}} |\braket{\psi}{\phi_{\rm prod}}|.
\end{equation}
The larger is $\Lambda_{\rm max}$ the less entangled is $\ket\psi$ because it is closer to a product state. Also, if $\ket\psi$ is the product of two entangled states, then the overlap will be the product of the maximum overlaps of each entangled state with its closest product approximation. Therefore it makes sense to define the geometric entanglement as an extensive function given by:
\begin{equation}
E_G= -\log_2 \Lambda_{\rm max}^2. 
\end{equation}
Notice that $E_G(\ket{\Phi^B})=1$ for a pair of qubits in a Bell state $\ket{\Phi^B}$ \cite{WeiPRA2005}. For the analysis of many-body systems composed of $N$ parties, it is also convenient to define the density of geometric entanglement or the geometric entanglement per party given by $\mathcal E= E_G/N$.
It helps our physical intuition to think of the state maximising the overlap as the mean-field approximation to a Hamiltonian having $\ket\psi$ as ground state, for example $H=-\ket\psi\bra\psi$.

Geometric entanglement has been applied in various 1D models close to quantum phase transitions. The scaling of GE of a region of size $\ell$ sites in an infinite translational invariant lattice was studied in \cite{BoteroReznik2007,OrusPRL2008}. In Ref.~\cite{BoteroReznik2007}, for a 1D chain of quantum harmonic oscillators and for a 1D XY spin-1/2 chain, Botero and Reznik found numerically that the density of GE increases logarithmically with the block size until it saturates when $\ell\sim\xi$, where $\xi$ is the correlation length of the system. 
For a critical system, $\xi\to\infty$, the density of GE is found numerically to diverge as:
\begin{equation}
\mathcal E = \frac{\kappa}{12} \log_2 \ell
\end{equation}
which is reminiscent of the scaling of the block entanglement entropy of critical 1D systems. Indeed Botero and Reznik found numerically, for both the bosonic chain and the XY model, that $\kappa\approx c$, where $c$ is the central charge of the conformal field theory associated to the corresponding quantum phase transition~\cite{BoteroReznik2007}.  

Immediately after, Or\'us \cite{OrusPRL2008}, showed, using conformal field theory, that indeed  geometric entanglement displays logarithmic scaling with the block size under successive renormalisation group transformations. Moreover he found a lower bound on the density of GE that indeed depends on the central charge:
\begin{equation}
\mathcal E < \frac{c}{6} \log_2 \ell
\end{equation}
thus confirming the results of Botero and Reznick. 

For finite systems at criticality, it was first found numerically using MPS that the density of GE scales with the total system size $N$ as~\cite{ShiNJP2010}:
\begin{equation}
\mathcal E = \mathcal E_\infty+ \frac{b}{N}+O(N^{-2}) 
\end{equation}
This scaling was confirmed for two critical spin-1/2 chains: the Ising model at the critical transverse field and the XXZ in the critical region (anisotropy $\Delta\in[-1;1]$). For the latter model, an analytical formula for the coefficient $b$ as a function of the anisotropy $\Delta$ was found in Ref.~\cite{StephanPRB2010}. Using conformal field theory, the authors of  Ref.~\cite{StephanPRB2010} found that the coefficient $b$ is related to a boundary term of the corresponding classical model:
\begin{equation}
b(\Delta)=-\frac{2}{\ln 2} s
\end{equation}
where the constant $s$ for critical models is universal and is equal to the Affleck-Ludwig boundary entropy associated to a boundary conformal field theory with a Neumann boundary. In terms of the anisotropy $\Delta$, the term $b$ is thus:
\begin{equation}
b(\Delta)=1-\log_2 R(\Delta)
\end{equation}
where $R(\Delta)=\sqrt{2/\pi \arccos\Delta}$ is the so-called compactification radius, related to the decay exponents of the correlation functions and to the Luttinger parameter.

\begin{figure}[t]
\begin{center}
\includegraphics[width=0.9\columnwidth]{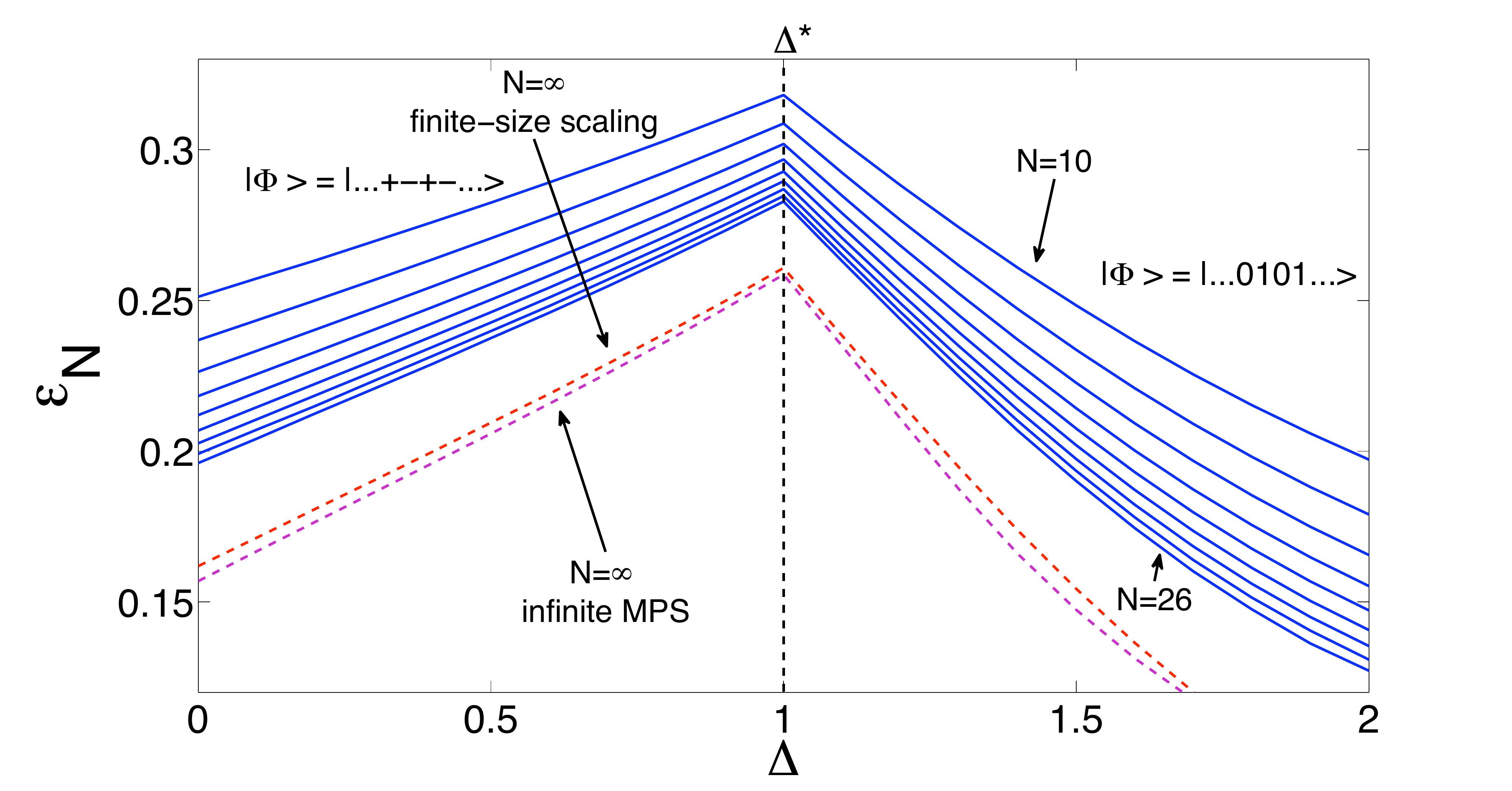}
\caption{Density of geometric entanglement as a function of the spin anisotropy $\Delta$. The BKT transition is indicated by a dashed vertical lines. The solid blue lines represent results obtained for finite system sizes for $10\le N \le 26$. The red dashed line represents the finite-size scaling extrapolation and the red purple line is the result of an infinite MPS calculation. (Results reprinted from \cite{OrusWeiPRB2010}).
}
\label{fig:GEBKT}
\end{center}
\end{figure}

The advantage in the use of GE is, however, more evident in the detection of more elusive quantum phase transitions that do not fit the Landau's paradigm. This is the case of the BKT transition, occurring for example in the critical to N\'eel transition in the spin-1/2 XXZ chain (see Sec.~\ref{sec:gallery}). This transition is sometimes called of infinite order because the ground state energy, its derivatives and the two-body correlation functions are all analytic at the critical point. Thus, entanglement measures based on two-body correlations, for example the concurrence, are unable to detect such a transition. Or\'us and Wei \cite{OrusWeiPRB2010} showed instead that the geometric entanglement exhibits a distinctive cusp at the BKT transition as shown in Fig.~\ref{fig:GEBKT}. The origin of the cusp is the following. Although the ground state is continuous across the transition, the closest product state that maximises the overlap changes abruptly for $\Delta=1$. For $\Delta<1$ the closest product state is $\ket{\Phi_{\rm prod}} = \ket{\dots +-+- \dots}$ where $\ket{\pm}$ are the eigenstates of $\sigma_x$. For $\Delta>1$, instead, the state that maximises the overlap is the N\'eel states $\ket{\dots 0101\dots}$, where $\ket0$ and $\ket 1$ are the eigenstates of $\sigma_z$. { We mention that recently a localisable geometric measure of entanglement has been employed to detect the BKT transition~\cite{SadhukhanPRA2017}.}

Recently, the geometric entanglement in the 1D cluster-Ising model has been studied close to the transition between an antiferromagnetic phase and a symmetrically protected topologically ordered phase, the so-called cluster phase whose ground state is adiabatically connected to a multipartite cluster state \cite{Wei2017}. 
Also, a curious connection between the geometric entanglement and the geometric phase has been established in Ref.~\cite{MousolouPRA2013}. Indeed geometric entanglement emerges as the real part of a complex observable quantity that can potentially be measured in interferometry experiments. 

The effect of temperature on geometric entanglement has been first discussed in Ref.~\cite{NakataPRA2009}. Technically, finding the GE for mixed states is much more difficult than for pure states because one should  perform first a minimisation over all possible (infinite) pure states decomposition of the mixed state at hand. The authors of Ref.~\cite{NakataPRA2009} overcome this problem using a lower bound on the geometric entanglement given by the robustness of entanglement (see Sec.~\ref{sec:bipartite}). This in turn gives a threshold temperature below which the state is certainly entangled\footnote{Because of the nature of the bound, it might be possible that a state above the threshold temperature is still entangled.}. The authors then study the threshold temperature for the XY model showing that it becomes very sensitive close to the zero-temperature quantum phase transitions of the model. Therefore an experiment able to measure the GE could detect the location of quantum phase transitions, even at moderate high temperatures. 

We have discussed above for 1D critical systems how the scaling of the density of GE per block scales logarithmically with the block size. This is the same scaling (apart from a different prefactor) of the scaling of two other bipartite measures of entanglement, namely the entanglement entropy and the single-copy entanglement. However this relation between entanglement measures depends crucially on the dimensionality of the system. Indeed for an infinite-dimensional model this is not the case. Or\'us, Dusuel and Vidal made a thorough analysis of the scaling of the single-copy entanglement and geometric entanglement in the Lipkin-Meshkov-Glick (LMG) model \cite{OrusDusuelVidalPRL2008}. They showed numerically that the GE (and not its density) behaves equivalently to the entanglement entropy and the single-copy entanglement with a logarithmic divergence approaching criticality.

\begin{figure}[t]
\begin{center}
\includegraphics[width=0.8\columnwidth]{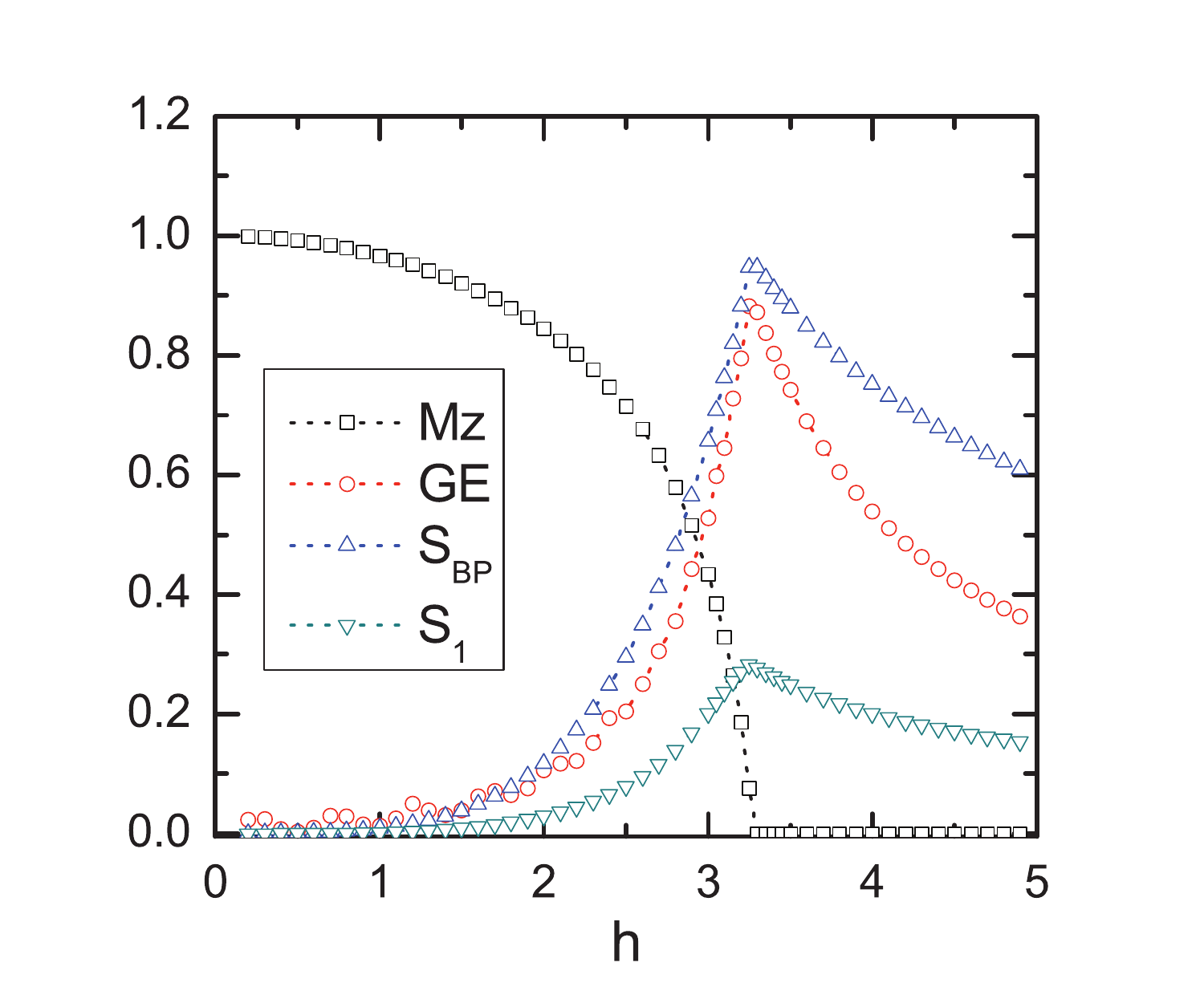}
\caption{Geometric entanglement and other measures for the 2D Ising model as a function of the transverse field. Geometric entanglement (circles), entanglement entropy (triangles up), single-site entropy (triangles down) and transverse magnetisation (squares) are calculated for a system of size $2^5\times 2^5$ using tensor product states.  (Results reprinted from \cite{HuangPRA2010}).
}
\label{fig:GE2DIsing}
\end{center}
\end{figure}

In 2D lattice systems the behaviour of geometric entanglement is not as well understood as for 1D systems. In Ref.~\cite{HuangPRA2010}, Huang and Lin, supported by numerical simulations involving matrix and tensor product states algorithms, analysed the GE for different 2D models: the transverse-field Ising model,  the XYX and the XXZ models. For a finite-size system, the GE, as well as the entanglement entropy and the single-site entropy (1-tangle), exhibits a sharp cusp at the critical point. For the Ising model the results are shown in Fig.~\ref{fig:GE2DIsing}. The derivatives of all quantities are discontinuous around $h\approx 3.25$ which is close to the expected location of the critical point, $h\approx 3.04$, as obtained by quantum Monte Carlo simulations.  The scaling of GE with the system size is briefly discussed in this paper and due to the exponential growth of computational resources needed, such analysis is unfeasible and is left as an open problem.

Recently Shi et al. studied GE in infinite square lattices using translational invariant infinite PEPS in several models in the class of Potts and Heisenberg models \cite{ShiPRA2016}. They showed that even for 2D quantum phase transitions associated to a symmetry breaking,  the behaviour of the GM is not conclusive. In \cite{ShiPRA2016} as well as in \cite{HuangPRA2010}, there are examples indicating that (i) there are discontinuous phase transitions (first order) in which the GM can be either continuous or not and (ii) there are continuous phase transitions in which is GE is continuous but its maximal value might not be at the critical transition point. 
These results therefore imply that the analysis of the GE alone is not sufficient to characterise completely the underlying quantum phase transition.

\begin{figure}[t]
\begin{center}
\includegraphics[width=0.7\columnwidth, trim={0 0 7.5cm 0},clip]{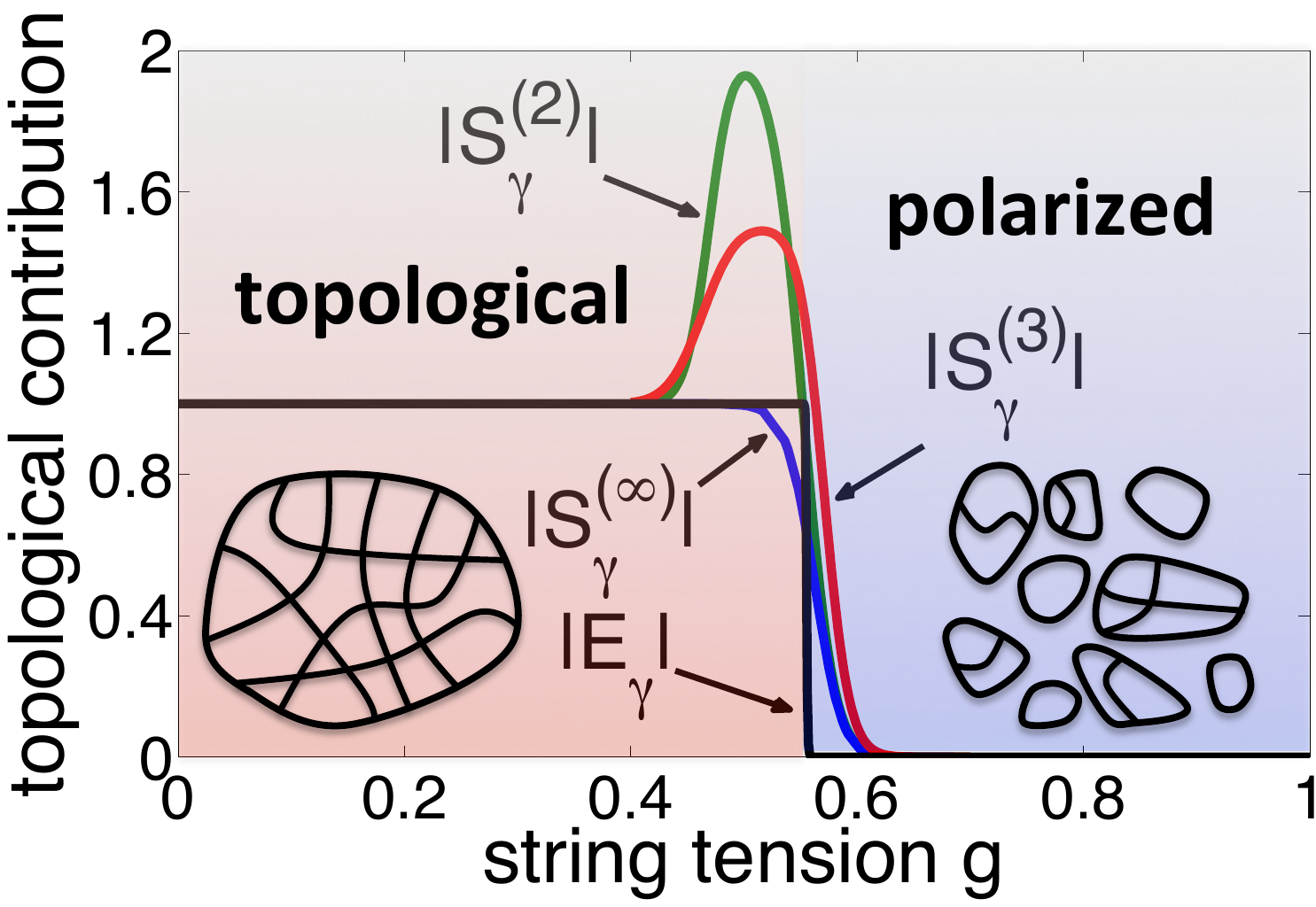}
\caption{Different topological entanglement measures for the toric code versus the string tension applied: GE (black), single-copy entanglement (blue) and Renyi entropies of order 2 and 3 (red and green, respectively). (Results reprinted from \cite{OrusPRL2014}).
}
\label{fig:GEtopological}
\end{center}
\end{figure}

For 2D systems another interesting feature that can emerge is topological order that does not fit the more traditional Landau's paradigm of symmetry breaking. The relationship between topological order and geometric entanglement was first analysed in Refs.~\cite{OrusNJP2014,OrusPRL2014}. In Ref.~\cite{OrusNJP2014}, Or\'us et al. analysed exactly solvable 2D models that exhibit topological order such as the toric code, double semion, color code and quantum double models. They show that, in analogy with the entanglement entropy, the GE of a block of size $L$ is a sum of two contributions: one, related to the bulk and proportional to $L$, thus fulfilling an area-law and a constant one that is non-zero only for systems with topological order~\cite{KitaevPRL2006}. In Ref.~\cite{OrusPRL2014}, the authors study numerically with the help of the PEPS algorithm the behaviour of the topological contributions to GE and bipartite Renyi entropies. Specifically, the authors consider the toric code with an applied string tension inducing a quantum phase transition between a polarised phase and a topological phase.
The results for the various entanglement measures are shown in Fig.~\ref{fig:GEtopological}. It is evident that, while all the Renyi entropies, including the single-copy entanglement, are smooth around the phase transition, the topological contribution to the GE remains approximately constant in the topological phase and drops sharply when entering the polarised phase. 

\subsection{Global entanglement}
A global measure of entanglement, originally defined by Meyer and Wallach \cite{MeyerWallach}, was adapted for systems of many interacting spins in \cite{Somma2004,deOliveiraPRA2006,deOliveiraPRL2006}. The global entanglement for $N$ qubits is defined as:
\begin{equation}
E_G^{(1)}=2-\frac 2N \sum_{j=1}^N {\rm Tr}(\rho_j^2) = \frac 1N \sum_{j=1}^N S_L(\rho_j) =\langle S_L\rangle
\end{equation}
where $\rho_j$ is the reduced density matrix of qubit $j$ and $S_L(\rho_j)=2[1- {\rm Tr}(\rho_j^2)]$ is the linear entropy, related to the purity of the single qubit density matrix. Thus, the global entanglement is the average linear entropy of all qubits. As such it is a collective measure of how much each qubit is entangled with the rest of the qubits. 
One advantage of the global entanglement is that it only involves single particle measurements aimed at estimating the purity of each qubit and could be thus easily assessed in experiments with quantum simulators of spin lattice systems with ultracold atoms or trapped ions.

Similarly to other measures of entanglement, global entanglement has generally a maximum at a second order phase transition. The global entanglement measure can be extended to include the two-qubit purity as follows \cite{deOliveiraPRA2006}:
\begin{equation}
E_G^{(2)}=\frac{4}{3(N-1)} \sum_{l=1}^{N-1}\left[1-\frac{1}{N-l}
\sum_{j=1}^{N-l} {\rm Tr}(\rho_{j,j+l}^2) \right ]
\end{equation}
where $\rho_{j,j+l}$ is the reduced density matrix of qubits $j$ and $j+l$. 
It is proven in Ref.~\cite{deOliveiraPRA2006} that $E_G^{(2)}$ better distinguishes different classes of multipartite entangled states {  of 2-level systems} like e.g., the N-qubit GHZ (Greenberger-Horne-Zeilinger) state of the form $1/\sqrt{2}(\ket{0}^{\otimes N}+\ket{1}^{\otimes N})$, the EPR (Einstein-Pololski-Rosen) state of the form 
$1/\sqrt{2}(\ket{01}+\ket{10})^{\otimes N}$ or W states.

Most applications of the global entanglement can be found in the general review \cite{AmicoRMP2008}. Although there have not been many developments for global entanglement since then---recent papers used mostly the geometric measure of entanglement and collective observables---very recently the global entanglement has been studied in the context of many-body localisation where it was used to pinpoint the location of the transition quite accurately~\cite{Filho2017}. We will discuss many-body localisation in Sec.~\ref{sec:otherquantum} in the context of quantum discord and quantum mutual information. 

\subsection{Genuine multipartite entanglement}
\label{sec:genuine}

So far we have discussed the entanglement of $N$ parties distinguishing them from product or separable states that do not contain any sort of entanglement.  The scenario of multipartite entanglement is however more complex, since there exist different classes of entangled states depending on how many parties are entangled across certain partitions. For example a pure state of three parties $A,B,C$ of the form:
\begin{equation}
\ket{\Psi}_{ABC} = \ket{\phi}_{AB}\otimes\ket\chi_C
\end{equation}
is in general entangled, as long as $\ket{\phi}_{AB}$ is entangled, however party $C$ is not correlated with $A$ and $B$ and in this sense state $\ket{\Psi}_{ABC}$ is not fully or genuinely multipartite entangled.
Here, we will not attempt to cover the full theory of genuine multipartite entanglement (GME), but in this section we will mention the latest developments in the study of GME for quantum many-body systems. We refer the reader to more specialised reviews on this subject, e.g. \cite{HorodeckiRMP2009,AmicoRMP2008,Eisert_multipartite_2016}.

It is useful for the following to introduce the notion of $k-$separability. A pure state of $N$ parties is called $k-$separable if it can be written as the product of at most $k\le N$ states $\ket\psi_i$ belonging to a set of non overlapping partitions $i=1,\dots k$:
\begin{equation}
\ket\psi = \ket\psi_1\otimes\ket\psi_2\otimes\dots\otimes\ket\psi_k.
\end{equation}
Thus $N-$separable states are simply product states and states that are not $k-$separable for any $k>1$ are fully entangled or GME. This definition can be extended to mixed states expressed as mixtures of states with certain $k-$separability. The notion of $k-$separability was first studied in spin chains in Ref.~\cite{GuhneNJP2005} in which an entanglement witness based on the ground state energy was devised. Specifically, for the Heisenberg model it was found that if the ground state energy $\bra{\psi_G} H\ket{\psi_G}$, where $\ket{\psi_G}$ is the ground state, is less than $-JN$ then the state is not $N-$separable, i.e. it is entangled. However if the energy is less than $-(1+\sqrt{5})/2 JN$ then the state is not 2-separable and contains some tripartite entanglement. A similar bound was found for the $XY$ model. 
 This kind of entanglement witnesses based on collective observables are particularly useful for experiments in which no single-particle addressing is possible. In Ref.~\cite{GabrielEPL2013}, Gabriel and Hiesmayr derived similar and more general bounds for $k-$separability and applied them to spin-1/2 and spin-1 chains. In Ref.~\cite{TroianiPRA2012}, Troiani and Siloi generalised the energy multipartite entanglement witness to clusters of arbitrary spin-s embedded in rotationally invariant spin chains.
The finite-size scaling of genuine multipartite entanglement was studied in various papers \cite{GiampaoloPRA2013,GiampaoloNJP2014,HofmannPRB2014,BayatPRL2017}.

In order to measure the genuine multipartite entanglement of pure many-body states it was proposed to use a generalised geometric measure (GGM) of entanglement in Refs.~\cite{SenDe2010,SenDePRA2010}.
Similarly to geometric entanglement the GGM of an N-party pure quantum state $\ket\psi$ is defined as:
\begin{equation}
\mathcal E_{GGM}(\ket\psi) = 1-\Lambda^2_{max}  
\end{equation}
where $\Lambda^2_{\max}=\max |\braket{\phi}{\psi}|$ where the maximisation is performed over the set of pure states $\ket\phi$ that are not genuinely multipartite entanglement, i.e., $\ket\phi$ is a product state for a certain partition. Notice that  $\ket\phi$ has some entanglement contrary to the fully factorised states used in the maximisation of the geometric entanglement. Thanks to the definition, a non-zero GGM implies the existence of genuine multipartite entanglement and excludes states that factorise across a particular bipartition. The maximisation in the definition of GGM is more complicated than the one needed for the GE. However it is possible to show that $\Lambda^2_{\max}$ is the largest Schmidt coefficient across all bipartitions and that the GGM is monotonically decreasing under LOCC. 
The GGM has been applied in the analysis of several models: the 1D $XY$ chains, Heisenberg and 2D frustrated models \cite{SenDe2010, SenDePRA2014,JindalPRA2014}, ladders \cite{SenDe2016}, and finally dimer (singlet) covering of square lattices and connection to spin liquids \cite{SenDePRL2013}. There exist more formal approaches to quantifying genuine multipartite correlations~\cite{BennettPRA2011,SzalayPRA2015,GirolamiPRL2017} and it would be interesting to apply these measures in strongly correlated systems.

Finally, a different approach was used in Ref.~\cite{StasinskaPRA2014} in which a characterisation of genuine multipartite rotationally invariant states was given equivalently to unitary invariant states \cite{EggelingPRA2001}. 
Rotationally invariant states of three qubits fulfil $\mathcal U\rho\mathcal U^\dagger = \rho$ where $\mathcal U=U\otimes U\otimes U$ and $U$ is an arbitrary unitary rotation of a qubit. They arise as ground and mixed states of rotationally invariant Hamiltonians. For qubits the only interaction with this feature is the Heisenberg one: $ \vec\sigma_i\cdot\vec\sigma_j$, as the dot product is invariant under a global axis rotation. Non degenerate ground states and thermal states of a rotationally invariant Hamiltonian are rotationally invariant. Moreover, the partial trace operation preserves this symmetry and hence, their reduced density matrices are also rotationally invariant.

\begin{figure}[t]
\begin{center}
\includegraphics[width=0.48\columnwidth]{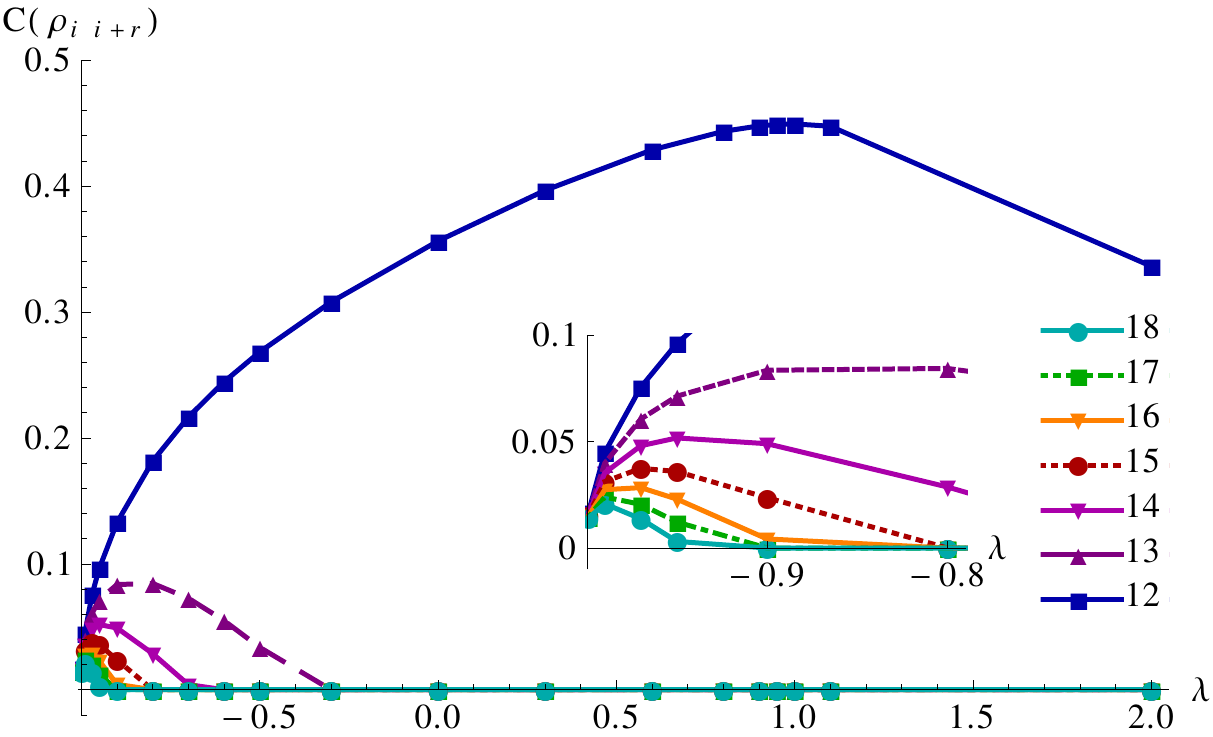}
\includegraphics[width=0.48\columnwidth]{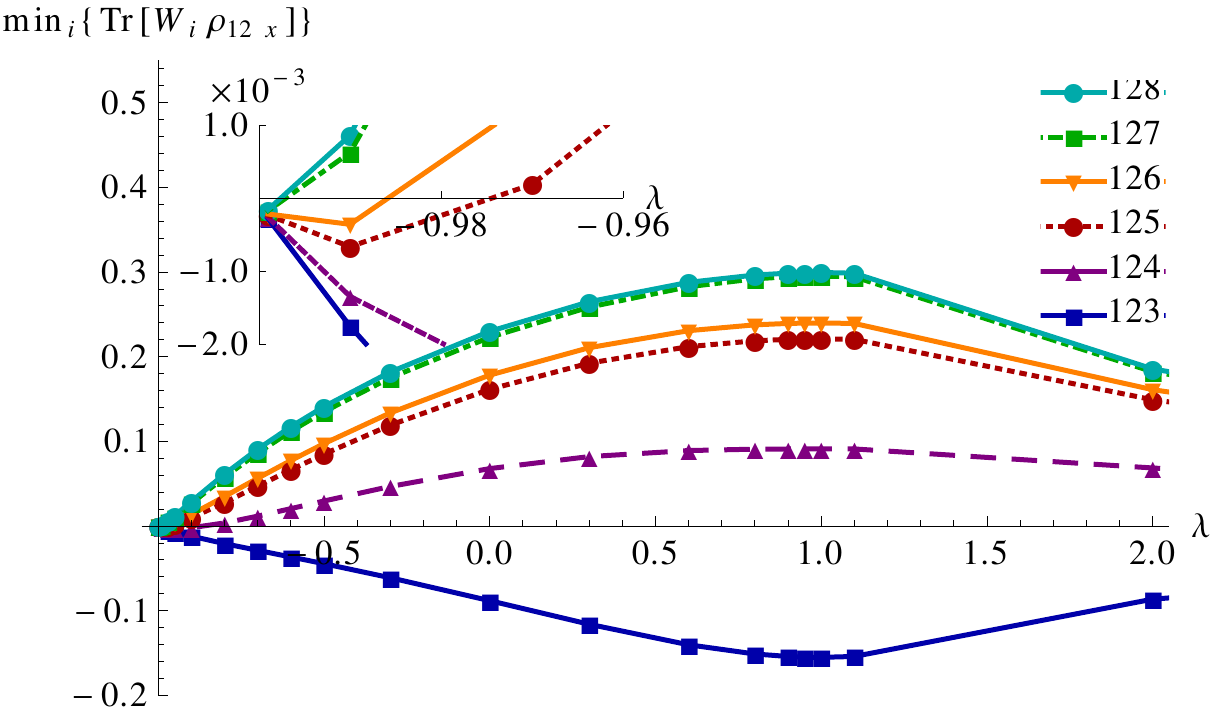}
\caption{Left: Concurrences of a reduced state of two non-adjacent spins $(i, i+r)$ at distance $r$ as a function of the anisotropy spin anisotropy, called $\lambda$ in these plots (we call it $\Delta$ in Eq.~\eqref{eq:Hmodel}). For convenience the spin arrangements
are denoted as $12, 13, \ldots$. The inset shows the results closer to the transition. Right: mean value of the witness detecting for negative values genuine tripartite entanglement of the reduced density matrix of three qubits. The legend describes the spin arrangements of the qubits. 
(Reprinted from Ref.~\cite{StasinskaPRA2014})}
\label{fig:Julia}
\end{center}
\end{figure}

The characterisation given in Ref.~\cite{StasinskaPRA2014} allows one to distinguish mixed separable, biseparable, and genuine multipartite entangled states of 3 qubits. 
This approach can also be extended to states that are not rotationally invariant by  applying a twirling map:
\begin{equation}
\Pi(\rho) = \int d\mathcal U\,\, \mathcal U\rho \mathcal U^\dagger
\end{equation}
where the integral is extended to all product of three unitaries: $\mathcal U=U\otimes U\otimes U$. Since the twirling map is  a non entangling operation, any entanglement detected in a twirled state existed in the original state. This approach was applied to the numerical study of the $XXZ$ spin-1/2 model near the first order transition between the ferromagnetic and critical phases and the results are shown in Fig.~\ref{fig:Julia}. As the transition is approached from the critical side, the concurrence of nearest-neighbour spins decreases while that of long-distance qubits increases. Simultaneously, genuine tripartite entanglement, detected by an entanglement witness, arises near the first order transitions between long-distance qubits.   



%% file: Chapter4.tex

\section{Quantum nonlocality in many-body systems}
\label{sec:nonlocality}

\subsection{Bell inequalities in many-body systems}

While the subject of entanglement in many-body systems have been throughly investigated in the last years, the characterisation of their nonlocal properties remains still largely unexplored and, only very recently, significant progress has been achieved \cite{Tura2014,Tura2015,TuraPRX2017}.

{  Nonlocality deals with the correlations between local independent observables in bipartite or multipartite states, see e.g. Ref.~\cite{Brunner2014} for a recent review on the subject. When the correlations cannot be reproduced by a local hidden variable (LHV) model, the state is said to be nonlocal. Such a restriction can  always be represented by a Bell inequality. Mathematically, a Bell inequality provides a bound on the strength of correlations between independent local observables that are compatible with a classical local hidden variable model or shared randomness.}

The detection of genuine quantum nonlocality for multipartite systems bears similitudes to the problem of multipartite entanglement. While it is well known that all pure entangled states are nonlocal~\cite{Gisin1991}, it is not known, except for a few cases, the relation between genuine multipartite entanglement and genuine multipartite nonlocality~\cite{Augusiak2015,ZukowskiPRL2002,ZukowskiPRA2002}.
 First rigorous inequalities detecting genuine multipartite nonlocality were introduced in Ref.~\cite{Svetlichny1987}, where Svetlichny analysed a 3-party system to go beyond bipartite nonlocality. He derived a new type of Bell inequality whose violation implies the presence of genuine tripartite nonlocality that, in turn, implies the existence of genuine tripartite entanglement.

To grasp the difficulty of constructing Bell inequalities for many-body systems, it is instructive to work in the device independent framework, in which Bell inequalities depend on the tuple $(N,m,d)$, where $N$ refers to the number of parties (spins, modes, \dots), $m$ the number of measurements available to each party, and $d$ the number of outputs associated to each measurement. Correlations between the different outputs resulting from a given set of performed measurements are given by the conditional probabilities:  $p(a_{k_1}^1\dots a_{k_i}^{i}\dots a_{k_N}^{N}|m_{k_1}^1\dots m_{k_i}^i\dots m_{k_N}^N)$, i.e., the probability that party $i$ when measuring observable $m_{k_i}$ obtains output $a_{k_i}$. All these probabilities are, of course, positive and must fulfil the conditions that $\sum_k p=1$, where $k$ is here a multi index corresponding to all possible outputs for each fixed  measurement set ${\cal{M}}=\{ m_1,\dots ,m_N \}$. In total, there are $(md)^N$ conditional probabilities fulfilling $p\ge 0$, and $m^N$ restrictions ($\sum_k p=1)$. Therefore, the number of independent conditions one should consider is $m^N(d^{N}-1)$, i.e., exponential in the number of parties! The above set of equations defines the polytope $\mathcal{P_S}$, that is, a convex set with a finite number of  extremal points (vertices) corresponding to the restrictions  $\sum_k p=1$.

A conditional probability distribution is said to be compatible with a LHV model if 
\begin{equation}
p(a_{1}\cdots a_{i}\cdots a_{N}| m_{0}\cdots m_{i}\cdots m_{N})
=\int_{\Lambda} d\lambda\; p(\lambda)\;\large\Pi_{i}\; p(a_{i} | m_{i},\lambda),.
\end{equation}
where $\lambda$ is a hidden variable, $p(\lambda)$ is a probability distribution of such variable, and $\Lambda$ is the space of hidden variables \footnote{To avoid burdening the notation we introduce the notation $\{a_1\dots a_i\dots a_N\}$ as a shorthand notation for $\{a_{k_1}^1\dots a_{k_i}^i\dots a_{k_N}^N\}$ and similarly for $\{m_{k_1}^1\dots m_{k_i}^i\dots m_{k_N}^N\}$.}. So, the correlations (global conditional probabilities) can be reproduced if the parties had access to a shared randomness $p(\lambda)$ known beforehand. Again, this set of probabilities form a polytope $\mathcal{L}$ whose vertices corresponds to the states for which all probabilities factorise:  
$p(a_1,\dots,a_N|m_1,\dots,m_N)=\Pi_{i=1}^{N}p(a_i|m_i)$ for all measurement sets $\{\mathcal{M}\}$ and outputs.

For a generic multipartite quantum systems, described by a density matrix $\rho_{1,\dots,N}$, the most general measurement is described by a positive operator valued measure (POVM), so that the measurement performed by party $i$ is described by a set of positive operators $\{\bf{\Pi}_i\}$, such that $\sum_i{\bf\Pi_{i}}=\id_{d}$. 
Conditional probabilities are calculated according to the Born's rule 
\begin{equation}
p(a_{1}\dots a_{N}| m_{1}\dots m_{N})=\Tr({\rho_{1,\dots,N}\;\;  {}\bf{\Pi}}^{m_{1}}_{a_{1}}\dots{\bf{\Pi}}^{m_{N}}_{a_{N}}).
\end{equation} 
States fulfilling such conditional probabilities form a convex set (i.e. infinite vertices) denoted by $\mathcal{Q}$. Imposing on the conditional probabilities further the non-signalling condition, i.e., that the choice of measurement $m_i$ made by party $i$ cannot be influenced by the choices made by the other $N-1$ parties,  

\begin{eqnarray}
\sum_{a_{i}} p(a_{1},\dots a_{i},\dots a_{N}|m_{1}\dots m_{i}\dots m_{n})=\\
=\sum_{a_{i}} p(a_{1},\dots a_{i},\dots a_{N}|m_{1}\dots m'_{i}\dots m_{N}),
\end{eqnarray}
Popescu and Rohrlich showed that this set, which is again a polytope $\mathcal{P}_{NS}$, is strictly larger than the convex set $\mathcal{Q}$  \cite{Popescu1994}. Therefore, according to the above discussion physical states are ordered along the following hierarchy: 
\begin{equation}
\mathcal{L}\subset \mathcal{Q} \subset \mathcal {P}_{NS} \subset \mathcal{P}_S
\end{equation}

\begin{figure}[h]
		\centering
		\includegraphics[width=0.6\textwidth]{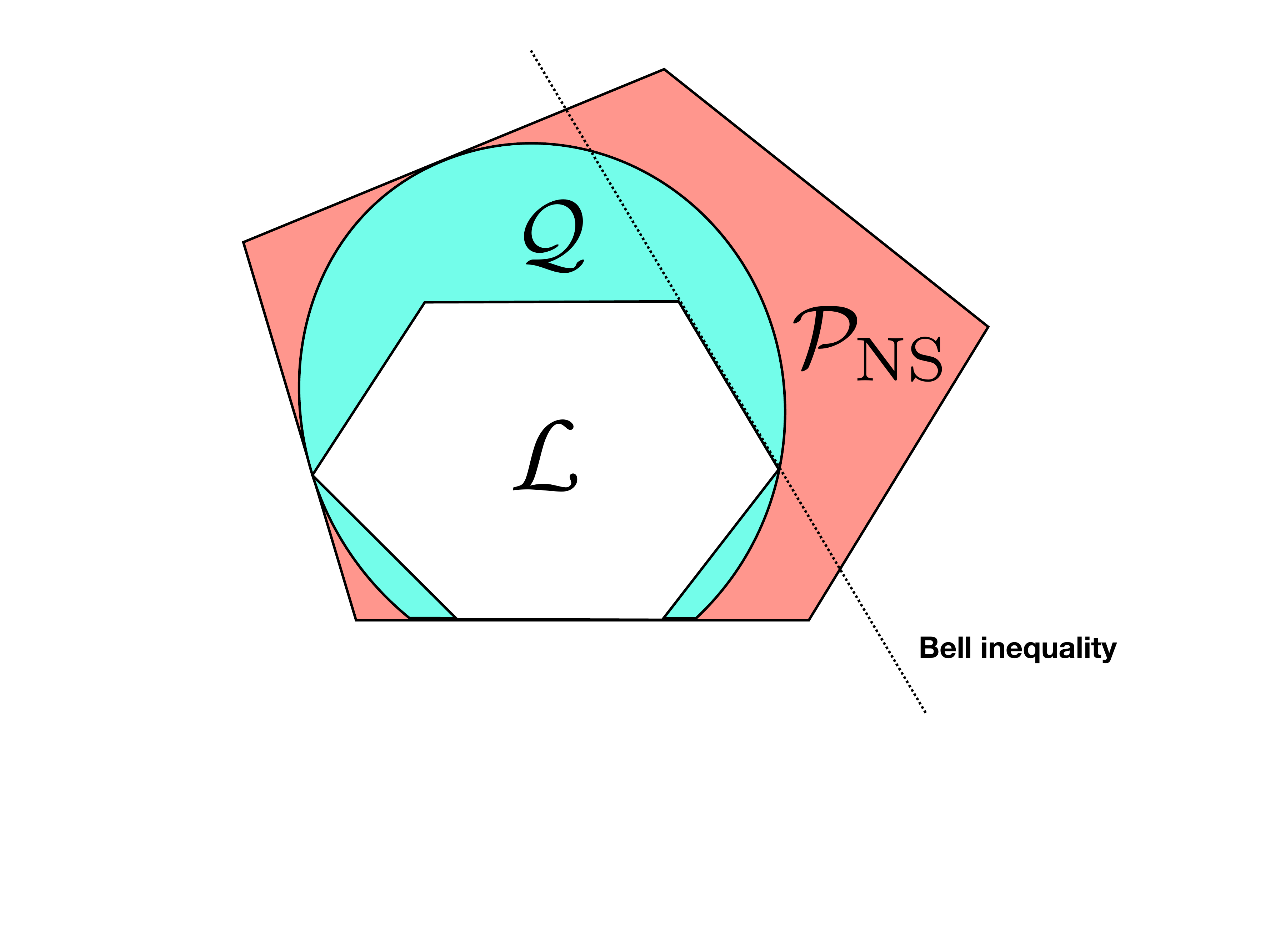}
		\caption{Sketch of correlations hierarchy: the inner set is the polytope $\mathcal L$ of local states, $\mathcal Q$ is the convex set of quantum nonlocal states, and $\mathcal P_{\rm NS}$ is the polytope of non signalling probabilities. Bell inequalities (dotted line) are hyperplanes delimiting the set $\mathcal L$ of local states.  }
		\label{fig:PPT_region}
	\end{figure}

A Bell inequality provides a test to check if a given state $\rho\in \mathcal{Q/L}$ or equivalently is the characterisation of the facets of the local polytope, $\mathcal{L}$, as schematically depicted in Fig.~\ref{fig:PPT_region}.

Although all {  pure} entangled states of many-body Hamiltonians are nonlocal~\cite{Gisin1991}, since constructing Bell inequalities requires to take into account all conditional probabilities, to characterise or even check the nonlocality of such ground states is an NP-hard problem. 
Since to construct a Bell inequality requires to take into account all conditional probabilities, to characterise or even to check if the ground or  an excited state of a many-body Hamiltonian is nonlocal is, in general, a NP-hard problem.

\subsection{Two-qubit nonlocality}

First studies of nonlocality in ground states of many-body systems were performed in the XY chain~\cite{Batle2010,Campbell2010} (see Eq.\eqref{eq:Hmodel} with $\Delta=0$). To this aim, the authors analysed 2-body Bell inequalities on the reduced density matrix of any two qubits located at a distance $R=|i-j|$. Due to the translational invariance of the system, the reduced density matrix reads~\cite{Barouch1970,Barouch1971}
\begin{equation}
	\rho_{ij}^{R}= \frac{1}{4}\left(\id + 
	\sum_{\mu,\nu=x,y,z}T_{uv}^{R}\sigma^{i}_{\mu}\otimes\sigma^{j}_{\nu}\right),
\end{equation}
with $T^{R}_{\mu\nu}= \langle \sigma_{\mu}^{i}\sigma_{\nu}^j\rangle$.
For symmetry reasons only the elements $\{T_{xx},T_{yy},T_{zz},T_{xy}\}$ are non vanishing. In the (2,2,2) setup, the simplest non trivial Bell inequality is the well known Clauser-Horne-Shimony-Holt (CHSH) inequality~\cite{CHSH}: 
\begin{equation}
\label{eq:CHSH}
\langle A_1 \otimes B_1 + A_1 \otimes B_2 + A_2 \otimes B_1 - A_2 \otimes B_2\rangle \le 2,
\end{equation}
where $A_i=\vec{a_i}\cdot\vec{\sigma}$ and $B_i=\vec{b_i}\cdot\vec{\sigma}$ being $\vec{a_i},\vec{b_i}$ unitary vectors in $\mathbb{R}^3$ {  and $\sigma$ the Pauli matrices}. Any probability distribution in $\mathcal L$ fulfils the above inequality but for example all maximally entangled states of two qubits violate it.  

In \cite{Batle2010,Campbell2010} it was shown that the CHSH inequality is never violated for such a model not even when the nearest-neighbour concurrence  verifies $C_{i,i+1}>0$, ensuring that there exists two-body entanglement. The same occurs for other marginals, like the 3-body reduced density matrix tested with Mermin-Bell inequalities~\cite{Mermin1990}. This result was generalised in \cite{RdeOliveira2012}, demonstrating that for any translational invariant spin chain Hamiltonian with two-body interactions, the two body reduced density matrices do not violate  Bell inequalities. It is worth mentioning that despite the fact that the CHSH inequality is never violated, the left-hand side of Eq.~\eqref{eq:CHSH} attains a global maximum 
at the quantum phase transition 
between the ferromagnetic and paramagnetic phases of the XY model. 
In Ref. \cite{Campbell2010}, multipartite Bell inequalities of the Svetlichny type~\cite{Collins2002,Svetlichny1987} were tested for an Ising ring up to $n=5$ spins showing that both ground state and low temperature Gibbs states violate them. For $n>7$ the formulation of such inequalities becomes unfeasible.

\subsection{Bell inequalities with one and two body correlators} 
A big leap in analysing multipartite Bell inequalities for the ground state of many body systems as compared to analyse Bell inequalities for the marginals was made in~\cite{Tura2014} by realising that one can simplify the polytope under study, $\mathcal{L}$, by imposing symmetries. For the interested reader we refer to \cite{Tura2015,ZizhuPRL2017} for a detailed explanation of these genuine many-body Bell inequalities of the type $(N,2,2)$ that can be applied to large multipartite $N$ systems,  opening thus a path in the study of nonlocality in quantum matter. A natural symmetry group is the permutational one, formed by states $\rho_{\Pi}$ which are permutationally invariant, i.e. they fulfil that $V_\sigma\rho_{\Pi}V_{\sigma}^{\dagger}=\rho_{\Pi}$, where $V_\sigma$ is the operator representing the permutation $\sigma$ over the $N$-element set. 
 It is possible to construct Bell inequalities that are invariant under such a group and most importantly that can be expressed using only one and two-body correlators~\cite{Tura2014}. This simplification has permitted to certify experimentally violation of genuine multipartite Bell inequalities in {  squeezed} BEC states~\cite{Schmied2016}. In \cite{WagnerPRL2017}, the authors introduced a new class of multipartite Bell inequalities involving two-body correlations but with an arbitrary number of measurement settings. If collective measurements are assumed, these new inequalities allow to address Bell correlations in many-body systems and to detect nonlocality in states that were not detected in
 \cite{Schmied2016}. Also preliminary studies investigating the nonlocal multipartite character between two spatially separated entangled BEC using~\cite{Tura2014} have been recently presented ~\cite{Lange2017}.

In the $d=2$ case, that is, when there is only two outputs for each measurement (for simplicity mapped to $\pm 1$), the Bell inequalities can indistinctly be formulated in terms of probabilities, or in terms of the expectation values of correlators where  $\langle\mathcal{M}_{j_{k}}^{i_{k}}\rangle=\pm 1$. Factorisation of probabilities means 
$\langle\mathcal{M}_{j_{1}}^{i_{1}}\dots \mathcal{M}_{j_{k}}^{i_{k}}\rangle=\langle\mathcal{M}_{j_{1}}^{i_{1}}\rangle \dots\langle\mathcal{M}_{j_{k}}^{i_{k}}\rangle$ for $k\le N$.  
As shown in \cite{Tura2014}, in a permutationally invariant set with $(N,2,2)$, in order to find the
maximal quantum violation of a Bell inequality with two dichotomic measurements per party, it is enough to consider qubits and traceless real observables with just one and two correlators. With all tools at hand, the most general Bell inequality 
bounding the local set in the permutational invariant manifold in the $(N,2,2)$ scenario is given by \cite{Tura2014}:

\begin{equation} 
\label{Bell}
\mathcal{B}(\theta , \phi) \equiv \alpha S_0 + \beta S_1 + \frac{\gamma}{2} S_{00} +\delta S_{01} + \frac{\epsilon}{2} S_{11} + \beta_C \geq 0,
\end{equation}
with
\begin{eqnarray}
S_l &\equiv & \sum\nolimits_{i=1}^N \left\langle  \mathcal{M}_l^{(i)} \right\rangle\\
S_{lr} &\equiv & \sum\nolimits_{i \neq j = 1}^N \left\langle  \mathcal{M}_l^{(i)} \mathcal{M}_r^{(j)}  \right\rangle,
\end{eqnarray}
for $l,r=0,1$,  where $\mathcal{M}_0 = \cos \phi \,\sigma_z + \sin \phi\, \sigma_x$, and  $\mathcal{M}_1 = \cos \theta\, \sigma_z + \sin \theta\, \sigma_x$, are the measurements. Here $\theta$, $\phi$ are the orientation angles of the measuring devices. In particular, Eq.~\eqref{Bell}, is violated by all entangled Dicke states of $N$-qubits $\ket{D^N_k} =(C^N_k)^{-1/2} \sum_{\sigma}\ket{\sigma(1^k 0^{N-k})}$, (i.e., $\forall k\neq 0,N$ ) for the following set of specific parameters \cite{Tura2014,Tura2015}:
$\nu = \left\lfloor {\frac{n}{2}} \right\rfloor - k$,
$\alpha = 2 \nu n (n-1)$,
$\beta = \alpha / n$,
$\gamma = n(n-1)$,
$\delta = n$,
$\epsilon= -2$, and
$\beta_C = {{n} \choose{2}} \left[ n+2(2 \nu^2 + 1) \right]$.
Defining  $Q(\ket{\psi}):=\bra{\psi} \mathcal{B} \ket{\psi}$, if $Q(\ket{\psi}) < 0 $, the state $\ket{\psi}$ violates the Bell inequality given by Eq.~\eqref{Bell}, and thereby the state $\ket{\psi}$ is entangled and genuine multipartite nonlocal due to its permutationally invariance. In \cite{Quesada2017}, it was 
shown that there is a large family of  even simpler states, the so-called Dicke diagonal symmetric, of $N$ qubits of the form $\rho=\sum_l c_k\ket{D^N_k}\bra{D^N_k}$ with $c_k>0$ ($\sum_k c_k=1$), that violate the weak Peres conjecture\cite{Peres1999}: those states are { entangled and positive under partial transpose with respect to one partition (PPT-bound entangled)}, but nevertheless they violate the above family of two-body Bell inequalities. Eq.~\eqref{Bell} also applies to the Lipkin-Meshkov-Glick spin model (see Eq.\eqref{eq:LMG}) whose ground state is precisely a Dicke state. Also, in Ref.~\cite{Pelisson2016}, the authors proposed to use the above Bell inequality, for an arbitrarily large number of neutral atoms trapped in a homogeneous one-dimensional optical  lattice, using the coupling  between  hyperfine  states  of  atoms  in  neighbouring wells, providing, thus, a new scenario where such inequalities could be used. 

\subsection{Nonlocality in translationally invariant spin Hamiltonians} 

In Ref.~\cite{TuraPRX2017}, the authors use dynamical programming, a classical optimisation technique based on backward induction, to determine optimal values a given multipartite local hidden variable model can attain for a given Hamiltonian. This value is used to construct Bell inequalities of the form $\id+\beta_c \le0$ with $\beta_c=-\min_{\rm LHV} I$ and Bell operators $\mathcal{B}=\beta_c\id +\mathcal{H}$ for some translational invariance spin chain Hamiltonians $\mathcal{H}$. The Bell inequality can be expressed through correlators of order $k$, $\langle\mathcal{M}_i\dots\mathcal{M}_{i+k}\rangle$ corresponding to a device independent scenario of $(N,m,2)$. Interestingly enough, such a construction allows one to check Bell inequalities by computing the energy of the Hamiltonian ground state and works for several one-dimensional spin model Hamiltonians, even for those that cannot be solved via the Jordan-Wigner transformation. In such cases, if the ground state energy is beyond the classical bound, it signals the presence of nonlocal correlations in the ground state.

\subsection{Quantum nonlocality in time: Leggett-Garg inequalities}
What distinguishes the classical macroscopic world from the quantum microscopic one? In quantum mechanics the act of measurement plays a fundamental role. In a classical system we assume that physical properties exist before the measurement takes place. In a more poetic way, the moon is there even if we do not look at it. Another property is that in classical mechanics we assume that the effect of a measurement can be arbitrarily reduced to zero. Using these two classical principles, dubbed macroscopic realism and noninvasive measurability, Leggett and Garg derived a series of inequalities that a theory must fulfil to satisfy the two principles \cite{LeggettGarg}. Any theory or experiment that violates any such inequality fails to satisfy one or both of those two principles. Thus, Leggett-Garg inequalities (LGI) provide a testable criterion for distinguishing the classical and the quantum world. 

The simplest LGI can be obtained by considering a dichotomic observable $Q$ that can assume the values $\pm 1$. We then measure its two-time correlation function $C_{ij} = \langle Q(t_i) Q(t_j)\rangle$ by performing correlations between measurements at three times $t_1,t_2,t_3$. If macroscopic realism and noninvasive measurability hold, then it must exist a probability distribution for the outcome of the three measurements. From the LGI, it follows that
\begin{equation}
K_3\equiv C_{21}+C_{32}-C_{31}\le 1.
\end{equation}
It is easy to see that a qubit performing Rabi oscillations violates the inequality. 
While Bell's inequalities probe the spatial nonlocality of two or more observations, LGI refer to the non-compatibility of two or more measurements in time. Thus LGI are often referred to as temporal Bell inequalities. For a complete review of LGI and their experimental tests see \cite{LGIreview}, here we focus to many-body systems.

In Ref.~\cite{GomezPRB2016}, G\'omez-Ruiz et al. use the LGI to detect the quantum phase transition occurring in 1D spin-chains. They consider the XXZ and anisotropic XY model (see Eq.~\eqref{eq:Hmodel}) and show that not only finite order transition can be detected by the LGI but also the infinite-order BKT transition. One of the advantages of the approach proposed by G\'omez-Ruiz et al., is that it is sufficient to monitor in real time the evolution of a local observable through the single-site two-time correlation:
\begin{equation}
\label{LGC}
C(t)=\frac 12 \bra{\psi_0} \{A(t),A(0)\}\ket{\psi_0}
\end{equation}
where $\ket{\psi_0}$ is the ground state of the Hamiltonian $H$ of interest, $\{\;\dots\;\}$ denotes the anticommutator and $A(t)=e^{iHt}A(0)e^{-iHt}$ is the evolution of the observable $A(0)$ in the Heisenberg picture.

The argument of G\'omez-Ruiz et al. is that, since $C(t)$ depends on the ground state energy and its derivatives with respect to the control parameter inducing the transition, then $C(t)$ should exhibit the same singularities occurring in the ground state energy and its derivatives. For example, in a first order transition one would expect $C(t)$, for a given time $t$  to be discontinuous as a function of the control parameter. As an example they apply their proposal to the XXZ chain, Eq.~\eqref{eq:Hmodel} with $\gamma=B=0$.
\begin{figure}[h]
	\centering
	\includegraphics[width=0.5\textwidth]{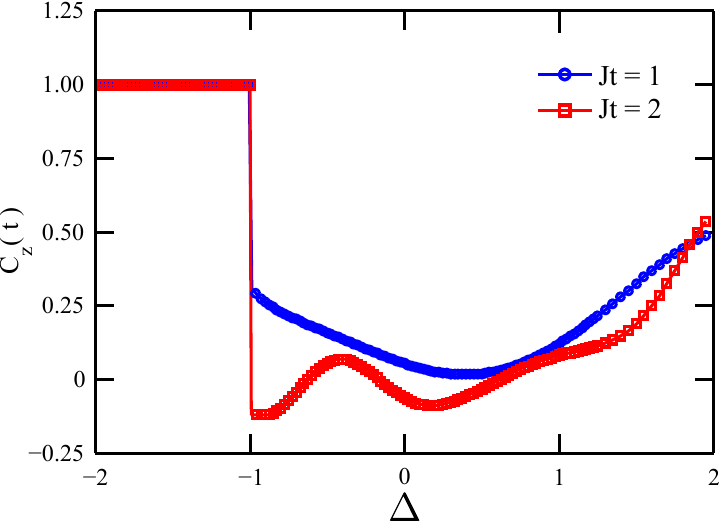}
	\caption{ $C_z(t)$ (see Eq.\eqref{LGC}), for two different times as a function of the anisotropy $\Delta$ in the XXZ spin chain. Results taken from \cite{GomezPRB2016}. }
	\label{fig:LG}
\end{figure}

In Fig.~\ref{fig:LG}, the quantity $C_z(t)$ for $A(0)=\sigma_z$ is shown as a function of $\Delta$ for two values of $t$. It is evident that $C_z(t)$ has a strong discontinuity at $\Delta=-1$ thus signalling the first order quantum phase transition. At $\Delta=1$, however, there is no clear trace of the BKT transition. In order to detect the latter, the authors of Ref.~\cite{GomezPRB2016} look at the maximal violation of the LGI using $A(0)=\sigma_x$ or $\sigma_z$. The authors also show that $K_3$ is always smaller than $-1$. However, for $|\Delta|>1$ the maximal violation comes from using $A(0)=\sigma_z$ while for $|\Delta|<1$ the maximal violation is attained for $A(0)=\sigma_x$. This behaviour leads to a discontinuity at $\Delta=-1$ and, surprisingly, a sharp peak, perhaps a cusp, at $\Delta=1$ thus signalling the BKT transition.

It is worth mentioning two other uses of the LGI in strongly-correlated systems. The first, in Ref.~\cite{LambertPRL2010}, deals with quantum transport through nano-structures. The authors distinguish classical and quantum transport using the violation of the LGI. This is strongly related to quantum non-Markovian and coherent effects occurring in the transport process. The second, very recent, considers the problem of macro-realism in two-well Bose-Einstein condensates~\cite{rosales2016leggett}. This idea, closer in spirit to Leggett and Garg's original paper, could be potentially well suited for experimental demonstrations in experiments with ultracold bosons in double wells.

%
%

%% file: Chapter5.tex

\section{Other quantum correlations}
\label{sec:otherquantum}


In this section  we consider other generalised forms of quantum correlations, including quantum discord, quantum mutual information and also turn our attention to quantum {coherence-based correlations}. Specialised reviews focussing more on quantum information aspects have recently appeared; we refer the interested readers to them \cite{RMPDiscord,AdessoQuantumCorrelations, streltsov2016quantum,AditiReview,Braun2017}. We instead concentrate on the application of such measures to the study of strongly correlated systems in one- and two-dimensional lattices.

\subsection{Bipartite discord}
Quantum discord has been introduced by {  Ollivier and Zurek} as an attempt to discriminate the correlations that characterise a bipartite system. It was soon realised that there exist separable quantum states, i.e. non entangled, that cannot be associated with classical states and that therefore contain some form of genuinely quantum correlations \cite{ZurekDiscord, VedralDiscord}. Quantum discord is defined as the difference of two quantities that for classical states, associated with a classical probability distribution, would vanish. The first quantity is the quantum mutual information:
\begin{equation}
I(\rho_{AB}) = S(\rho_A)+S(\rho_B)-S(\rho_{AB}),
\end{equation}
where $S(\rho)$ is the von Neumann entropy. The mutual information measures how much information is contained in the bipartite system $\rho_{AB}$ that cannot be accessed by looking at the reduced states $\rho_{A}$ and $\rho_B$. The second quantity is the one-way classical information and measures the amount of information that can be gained on the system $B$ if one performs measurements on $A$:
\begin{equation}
J(\rho_{AB}) = S(\rho_B)-S(\rho_{AB}|\Pi_A),
\end{equation}
where the measurement-based conditional entropy is:
\begin{equation}
S(\rho_{AB}|\Pi_A) = \sum_j p_j S(\rho_{B|j}),
\end{equation}
and is the average amount of information retrieved when measuring the system $A$ and obtaining, with probability $p_j$ the outcome $j$, collapsing $B$ in the  state $\rho_{B|j}$.
If $A$ and $B$ were classical random variables then $I$ and $J$ would be equal thanks to Bayes' rule. Quantum discord is precisely defined as the difference of these two quantities in a quantum setting:
\begin{equation}
\label{eq:definitiondiscord}
D(\rho_{AB}) = I(\rho_{AB})-\max_{\Pi_A} J(\rho_{AB}).
\end{equation}
Naturally, the amount of information retrieved depends on the correlations shared between $A$ and $B$ but also on the measurement performed. Therefore we need to maximise the one-way classical information over all possible measurements: projective von Neumann measurements as well as POVM that can be thought as projective measurements in an enlarged Hilbert space. Ref.~\cite{RMPDiscord} contains a more extended introduction to quantum discord and related measures in the quantum information context. 

The first papers that applied quantum discord to measure quantum correlations between two nearest-neighbouring spins in a one-dimensional spin chains were by Dillenschneider \cite{Dillenschneider2008} and by Sarandy \cite{SarandyPRA2009}. Dillenschneider   \cite{Dillenschneider2008} studied the ground state of the Ising and XXZ spin-1/2 chain. He showed that the first derivative of the discord diverges at the ferro-paramagnetic transition of the transverse-field Ising model. For the XXZ model, he showed that discord exhibits a cusp at the critical-N\'eel transition due to a level crossing of the ground and first excited states. In contrast the concurrence is insensitive to such a crossing and appears to be analytic even in the thermodynamic limit.
Sarandy \cite{SarandyPRA2009} extended these results and showed the scaling of the first derivative of the discord at the ferro-paramagnetic transition and applied the discord to the infinitely-connected LMG model (see Eq.\eqref{eq:LMG}) at zero temperature. Concurrence and quantum discord have also been computed for the fermionic bond-charge extended Hubbard model \cite{AllegraPRB2011}, a paradigmatic model in condensed matter physics. 

There have been some investigations of quantum discord also for two-dimensional lattices. In Ref.~\cite{ChenLiPRA2010}, Chen and Li studied quantum correlations in a deformed toric model \cite{CastelnovoChamon} which undergoes a quantum phase transition from a topologically ordered phase to a magnetised one. 
In contrast to 1D chains, for this model quantum discord is always zero for any pair of qubits. The authors analyse a global form of quantum correlations which clearly detects the critical point from a discontinuity in its derivative. This is probably a specific feature of this model and not a general feature of discord. Indeed quantum discord was calculated in Ref.~\cite{SellmannPRB2015} for the spin-1/2 XXZ model on the triangular lattice using the two-dimensional DMRG. Quantum discord and concurrences correctly detect the quantum phase transitions through various magnetic orderings. 

The advantage of discord is evident when considering thermal states for which two-qubit entanglement, e.g. concurrence, vanishes abruptly.  One of the first analysis of the quantum discord of two nearest neighbour spins in a Heisenberg spin chain  at non zero temperature was performed in Refs.~\cite{WerlangPRL2010,WerlangPRA2010}. Surprisingly, there it was shown that quantum discord can increase with temperature if low-lying excited states contain more quantum correlations than the ground state. Moreover, in contrast to the concurrence, quantum discord is non zero also for distant spins and stills detects a quantum phase transition  \cite{MazieroPRA2010}. The scaling of discord with the distance of the two spins has been recently calculated analytically in Ref.~\cite{huang2014scaling} for various integrable spin chains.   
Quantum discord at non-zero temperatures has also been discussed in relation to symmetry breaking in Ref.~\cite{TomaselloEPL2011} and  to factorisation in Ref.~\cite{CampbellPRA2013,canosa2017quantum}. For further details on the earlier works we refer the readers to ~\cite{ReviewSarandy,WerlangRibeiroRigolin}. 

Discord in random spin chains has also been addressed by Sadhukhan et al.~\cite{Sadhukhan2016a,Sadhukhan2016b}. The authors studied random anisotropic XYZ chains with quenched disorder in the couplings and in the local magnetic field. In particular, they analysed the decay of correlations between two spin-1/2 with their mutual distance for three models: an ordered quantum XY chain, a random-field quantum XY model and a quantum XY spin glass.  In fact, they found that  contrary to the concurrence, discord decays exponentially with distance with a characteristic discord length. In some cases, depending on the couplings, the discord length is significantly larger in the disordered system than in the clean model. 
It remains an open problem whether there exists a connection of the discord length with the more traditional correlation length.

The time evolution of two-qubits quantum discord was analysed in various works following an instantaneous change of a Hamiltonian parameter \cite{DharEPL2012,Mishra2013, Mishra2016,ChandaPRA2016} or when such a parameter is changed linearly in time \cite{NagJSTAT2011}. 
Contrary to the time evolution of entanglement, which can exhibit collapses and revivals, dramatically known in the literature as sudden deaths and births~\cite{YuPRL2004}, the dynamics of quantum discord is not affected by collapses.
In Ref.~\cite{DharEPL2012}, it was found that the dynamics of discord may signal the revival of entanglement after a collapse provided that the amount of discord is large enough. In Refs.~\cite{Mishra2013, Mishra2016,ChandaPRA2016}, it is shown that the dynamics of discord and entanglement are qualitatively different, including their long-time limit behaviour. Surprisingly, there are instances in which the entanglement is ergodic, i.e. the long-time limit or average is equal to the canonical value, but discord is not. The origin of this ergodic-to-nonergodic transition in information-theoretic measures of quantum correlations is still unclear. However its explanation could shed some light on the topic of thermalisation in closed systems. 

The effect of a critical environment on the two-qubit discord has been analysed in Ref.~\cite{XiJPB2011}, in which two qubits are individually coupled collectively to two Ising closed chains, reminiscent of earlier studies of the Loschmidt echo in critical environments~\cite{QuanZanardiPRL2006}. It is found that the decay of quantum discord shows a strong variation when the two Ising chains are tuned to their quantum critical point. The effect of a Markovian bath on the quantum discord of two-nearest neighbour spins in an XY spin chains has been analysed in Ref.~\cite{Pal2012}. 
In Ref.~\cite{JoshiPRA2013}, Joshi and co-workers analyse the non-equilibrium steady state of a driven-dissipative 1D interacting system. This is  made of an arrays of coupled cavities, each containing a non-linear medium inducing a non-linear term and driven by an external pump. In this non-linear regime, the dynamics of each cavity can be well approximated by retaining only the vacuum and one-photon state. Therefore the dynamics can be mapped onto a spin-1/2 XY chain in a transverse field subject to Markovian dissipation (due to the cavity losses). The authors look at negativity and quantum discord in the steady state computed with matrix product states. They show that in contrast to negativity which vanishes for small transverse fields, since the state is approaching a mixture of separable states, quantum discord remains finite. At the critical point both, negativity and discord, exhibit a maximum, however both quantities appear analytical. 

For two spin-1 particles, quantum discord is much more challenging to calculate. { Indeed calculating discord for increasing Hilbert space dimension is an NP-complete problem~\cite{HuangNJP2014}. See Ref.~\cite{AditiReview} for a general discussion of the computability of quantum discord.} The reason is that the optimisation in the definition of discord in Eq.~\eqref{eq:definitiondiscord} must be performed over all possible measurements. Restricting these to projective measurements, thus leading to an upper bound of discord, the maximisation for qubits consists in choosing a measurement basis of two orthogonal states corresponding to a direction in the Bloch sphere, parametrised by the two polar angles $\theta$ and $\phi$. For spin-1, instead, simply rotating a state with maximum angular momentum is not enough. Loosely speaking one should also introduce bases obtained by spin-squeezing transformations. The complete characterisation of the measurement optimisation procedure for the calculation of spin-1 discord was reported in \cite{RossignoliPRA2012}. There, Rossignoli, Matera and Canosa show that 6 angles are needed to parametrise the most general orthogonal measurement for a spin-1 particle and apply this to a few examples. An application for such a measurement for the global discord of spin-1 chains is described in Sec.~\ref{sec:globaldiscord}.

\subsubsection{Experimental measurements of quantum discord}
We conclude this section on quantum discord with a survey on the latest experimental measurements. The relaxation dynamics of two-qubit quantum discord has been measured in nuclear magnetic resonance (NMR) experiments \cite{AuccaisePRL2011} (see also Refs.~\cite{CeleriIJQI2011,PassantePRA2011,KatiyarPRA2012,SilvaPRL2013} for other recent experimental measurements of quantum discord in NMR setups). 

Solid state spin chains have allowed to observe long-distance entanglement as well as quantum discord~\cite{SahlingNatPhys2015}. The different forms of quantum correlations have been measured against temperature. Around the Curie temperature, entanglement suddenly disappears although discord is observed for higher temperatures. 

Recently, the dynamics of quantum discord between the spin state of an ion and the motional degrees of freedom of an ion crystal has been observed in  Ref.~\cite{abdelrahman2016local}. The experiment consists in preparing a 1D self-organised array of ions in a highly anisotropic Paul trap, i.e. a Coulomb crystal. Following an earlier proposal \cite{DeChiaraPRA2008}, a Ramsey scheme is implemented in which the internal electronic levels of one ion of the chain, are subjected to a $\pi/2$ pulse by an external laser and, the resulting photon recoil induces a dynamics in the ions spatial degrees of freedom. 
The effective motional decoherence induced on the internal levels of the ion 
permits to extract the normal frequencies of the crystal even at moderate temperature. Moreover, using the results of Ref.~\cite{GessnerPRA2013}, the authors can extract the amount of quantum discord between the internal levels of the illuminated ion and the motional degrees of freedoms of the Coulomb crystal in terms of the two-time auto-correlation function of the ion.

\subsection{Quantum correlations based on response functions}
A particularly original approach to quantum correlations which is not motivated by an information theoretical resource approach has been recently reported in Ref.~\cite{MalpettiRoscilde}, 
where quantum correlations are defined in terms of conventional correlation functions and response functions in statistical mechanics. Their definition of quantum correlation functions goes as follows. Let us assume a generic quantum system described by a Hamiltonian $H$ and in the equilibrium state $\rho=e^{-\beta H}/Z$ at inverse temperature $\beta=1/(k_BT)$ and with the partition function $Z={\rm Tr}e^{-\beta H}$. Let us consider two local observables $O_A$ and $O_B$ that have support in two spatially separated regions $A$ and $B$ of the system. The system is subject to a perturbation acting on system $B$ only and described by a Hamiltonian term $-\lambda_B O_B$.

The quantum correlation function (QCF) between $A$ and $B$ is defined as follows:
\begin{eqnarray}
\langle \delta O_A \delta O_B \rangle_Q  & = &  \langle \delta O_A \delta O_B \rangle - \frac{\partial\langle O_A \rangle}{\partial \lambda_B}\Big |_{\lambda_B=0}  \\
& = & \langle \delta O_A \delta O_B \rangle - \frac{1}{\beta} \int_0^{\beta} d\tau \langle \delta O_A(\tau) \delta O_B(0) \rangle~. \nonumber
\label{e.QCF}
\end{eqnarray}
where all averages $\langle \dots\rangle$ are taken with respect to the state $\rho$; $\delta O = O - \langle O \rangle$ is the fluctuation of an operator from its average value; and $O(\tau) = e^{\tau H} O e^{-\tau  H}$ is the imaginary-time-evolved operator.
The QCF measures the difference between the two point correlations $ \langle \delta O_A \delta O_B \rangle$ and the response of region $A$ to the perturbation proportional to $\lambda_B$ in region $B$. While in a classical system $\langle \delta O_A \delta O_B \rangle_Q$ would automatically vanish because of the fluctuation-dissipation relation, for a quantum system this quantity is non-trivial because of non-commutativity of operators $O_A$ and $O_B$ with the system Hamiltonian $H$. 

\begin{figure}[t]
\centering
\includegraphics[width=0.6\columnwidth]{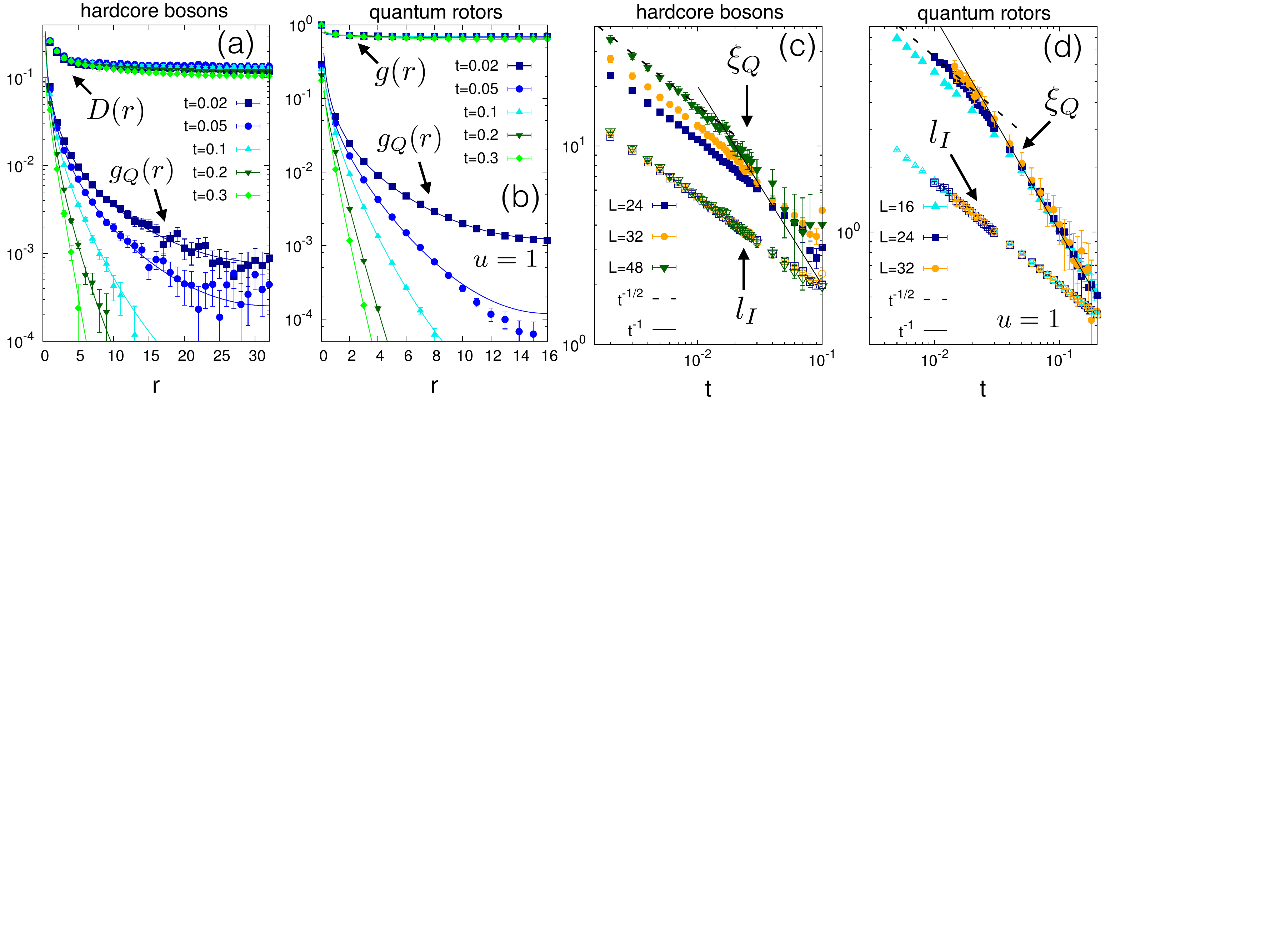}
\caption{Quantum correlation functions $g_Q(r)$ defined in the text compared to the standard two-point correlations function $g(r)$ and to the two-point quantum discord $D(r)$. Panel (a) is for hard-core bosons and panel (b) for quantum rotors. The parameter $t=k_BT/J$ is the temperature and $J$ is the tunnelling amplitude. 
From Ref.~\cite{MalpettiRoscilde}.
}
\label{fig:QCF}
\end{figure}

The authors of Ref.~\cite{MalpettiRoscilde} apply this measure to two paradigmatic examples: 1) hard-core bosons and 2) quantum rotors on a 2D square lattice. These models exhibit a transition from a normal phase at high temperatures and a superfluid phase for $T<T_{BKT}$ below the Berezinski-Kosterlitz-Thouless transition characterised by the divergence of the correlation length. It is important to stress at this point that the BKT transition is fully driven by thermal fluctuations that dominate over quantum fluctuations. Therefore one would not expect a genuine measure of quantum correlations to survive at this temperature or to have any non-analytic behaviour. 

For both models the QCF can be calculated numerically using quantum MonteCarlo techniques.
The results for the QCF $g_Q(r)=\langle a^\dagger_i a_{i+r} \rangle_Q$ for the creation and annihilation operators $a_{i\in A}$ and $a^\dagger_{j\in B}$ pertaining to two separated regions $A$ and $B$ are shown in Fig.~\ref{fig:QCF} as a function of the separation $r$ and for various temperatures. These are compared to the standard two-point correlation function $g(r)=\langle a^\dagger_i a_{i+r} \rangle$. $D(r)$ is the corresponding quantum discord of sites $i$ and $i+r$.

It is evident that quantum discord $D(r)$ and the standard correlation functions $g(r)$ are much larger and decay much slower than the QCF $g_Q(r)$.
In particular, one can show that quantum discord for the hard-core bosons decays algebraically in the superfluid phase and it is singular at the BKT transition even though the transition is believed to be driven only by thermal fluctuations. In contrast, the QCF is exponentially decaying throughout the superfluid and normal phases and does not show any signature of the BKT transition. From its decay one can extract a characteristic length $\xi_Q$, dubbed ``quantum coherence length" in \cite{MalpettiRoscilde}, which diverges as $T\to 0$ with a characteristic exponent. At the time of writing, it is still unclear whether this exponent is related to a quantum critical exponent or whether it is universal. In any case, the divergence of the quantum coherence length seems to suggest that the QCF is indeed a genuine measure of quantum correlations, at least from a physical point of view.
{ An intuitive reason why quantum discord is sensitive to a classical phase transition is because it is a function of the reduced density matrix of two sites. On the other hand, the QCF contains all the structure of quantum correlations in imaginary time and real space, thus, in a sense, contains more information than standard correlation functions.}

\subsection{Global discord in spin chains}
\label{sec:globaldiscord}
Global quantum discord (GQD) was introduced in \cite{RulliSarandy} as a multipartite extension of a symmetric version of bipartite discord. For two parties $A$ and $B$, 
 a symmetrised version of the QD can be obtained with a local measurement $\Pi_{ij} = \Pi_A^i\otimes \Pi_B^j$ such that $\Pi(\rho_{AB}) = \sum_{ij} \Pi_{ij}\rho_{AB}\Pi_{ij}$. We define the symmetric QD
\begin{equation}
\label{eq:discordsymm}
\mathcal{D}_2(\rho_{AB})=\min_{\{\Pi\}} \left\{S(\rho_{AB}||\Pi(\rho_{AB}))
-\sum_{\alpha=A,B}S(\rho_{\alpha}||\Pi^\alpha(\rho_\alpha))\right\},
\end{equation}
where $\rho_\alpha$ is the  corresponding reduced density matrix and $S(\rho||\sigma)$ is the relative entropy (Eq.\eqref{eq:relative_entropy}),
which vanishes as $\sigma$ approaches $\rho$.
Because the relative entropy is a measure of distinguishability between both states (albeit with all the caveats as it is not a distance since it is not symmetric), Eq.~\eqref{eq:discordsymm} can be interpreted as the difference between two terms. The first one, that is global on parties $A$ and $B$, represents the distinguishability between the original state, $\rho_{AB}$, and the state $\Pi(\rho_{AB})$ obtained after measuring locally the state $\rho_{AB}$. The second term is the sum of the relative entropies between the reduced states $\rho_\alpha$ and the locally measured one $\Pi^\alpha(\rho_\alpha)$, $\alpha=A,B$. Thus \eqref{eq:discordsymm} quantifies the discrepancy of a global measurement induced disturbance (although made with local observables) and the local measurement induced disturbances on each party. If the states are classical they are not disturbed and therefore all relative entropies vanish and $\mathcal{D}_2$ is zero. 

 In \cite{RulliSarandy}, Rulli and Sarandy showed that Eq.~\eqref{eq:discordsymm} can be generalised to multipartite states. For a quantum system comprising of $N$ subsystems in the state $\rho$, 
and assuming local projective measurements \footnote{Although the definition is valid for any generalised measurement, for simplicity we assume here local projective measurements.}  $\Pi(\rho)=\sum_{k} \Pi_k \rho \Pi_k$ where $\Pi_k  = \Pi^{k_1}_1 \otimes \Pi^{k_2}_2\otimes \dots \otimes\Pi^{k_N}_N$,  
the global quantum discord (GQD) is defined as:
\begin{equation}
\label{GQD}
\mathcal{D}_N(\rho)=\min_{\{\Pi\}}\bigg\{S\left(\rho||\Pi(\rho)\right)-\sum_{\alpha=1}^N S\left(\rho_{\alpha}||\Pi^\alpha(\rho_\alpha)\right)\bigg\}
\end{equation}
Again, this definition can be interpreted as the difference between the relative entropy of the whole state with its measured counterpart on one hand, and the sum of the relative entropies of the local states and their measured counterparts on the other. 
The minimisation in Eqs.~\eqref{eq:discordsymm}-\eqref{GQD} is, in principle, over all possible measurements, not only projective ones. However, because of the technical difficulties of characterising all possible POVMs, most of the literature, so far, has restricted the minimisation over projective measurements. Rulli and Sarandy showed that $\mathcal{D}_N(\rho)$ is able to detect, in contrast to entanglement and discord, the infinite-order quantum phase transition in the Ashkin-Teller chain \cite{RulliSarandy}.
In Ref.~\cite{Campbell2011}, the authors analyse $\mathcal{D}_N(\rho)$ in an Ising chain with periodic boundary conditions finding evidence of the ferro-paramagnetic phase transition in the peak of the GQD.

\begin{figure}[t]
\begin{center}
\includegraphics[width=0.5\columnwidth]{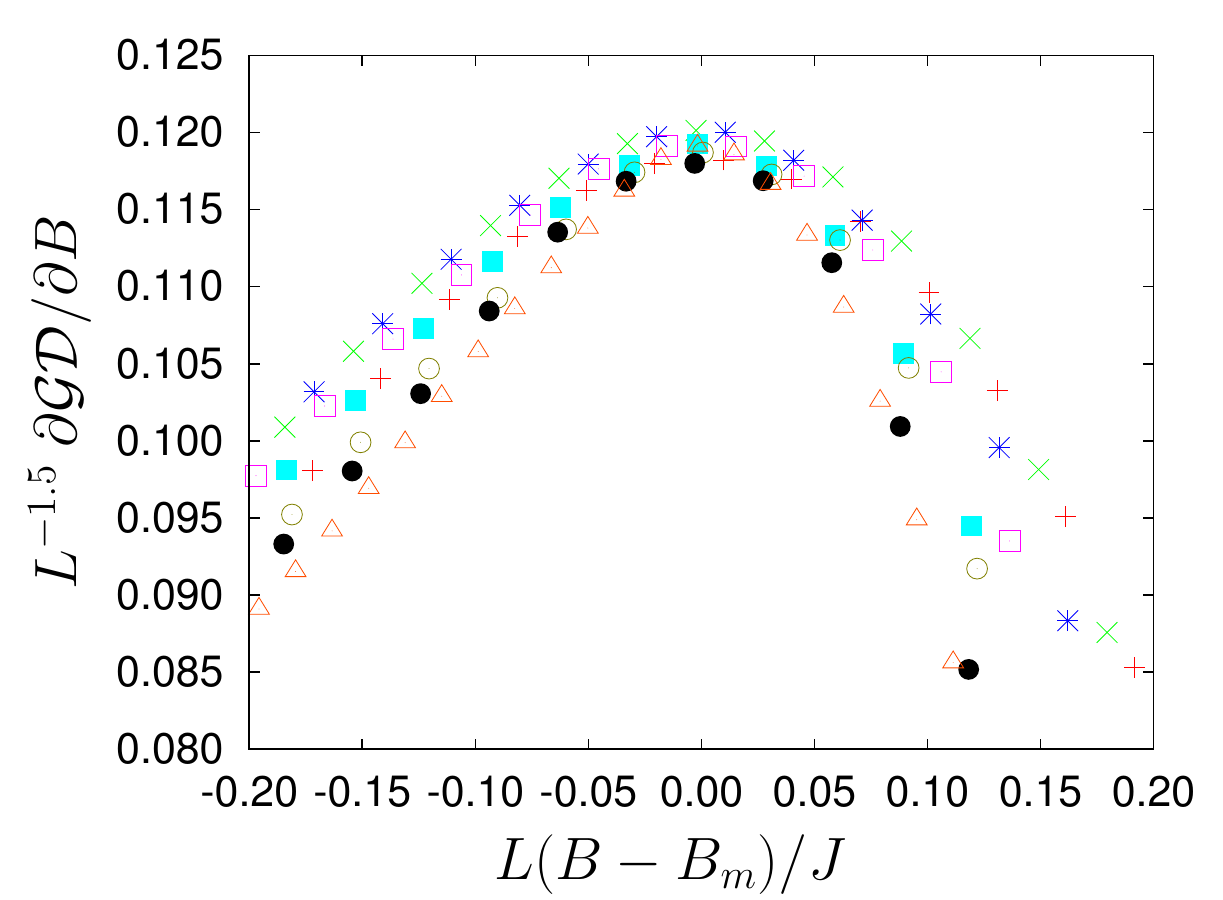}
\caption{Derivative of GQD with respect to the magnetic field for the Ising model at zero temperature. The data points (for $L = 4, . . . , 11$) are scaled according to the number of spins of the rings. $B_m$ is the critical value of the magnetic field for a finite-size chain of length $L$. Close to the critical point the curves at all values of L collapse to a common function, witnessing universality. (From Ref.~\cite{CampbellNJP2013})}
\label{fig:GQDising}
\end{center}
\end{figure}

Later it was shown that global quantum discord is not only able to detect the critical point of spin-Hamiltonians but follows universal scaling relations involving critical exponents. In Ref. \cite{CampbellNJP2013}, Campbell et al., studied the scaling of GQD in three spin-chain models: the transverse-field Ising chain, the cluster-Ising model \cite{SonEPL2011} and the XY chain with transverse field, both at zero and non-zero temperature.  
For the spin-1/2 Ising chain, the authors find that the first derivative of the GQD follows universal scaling:
\begin{equation}
\frac{\partial}{\partial B}{\cal D}_N=L^{-\omega}f[L^{\nu}(B-B_m)/J]
\end{equation}
where $B_m$ is the position of the maximum of the GQD as a function of the transverse field $B$. The exponent $\nu=1$, is the universal exponent of the correlation length in the  Ising universality class while $\omega\approx-1.5$ should be the critical exponent characterising the scaling of GQD for this transition. The results are plotted in Fig.~\ref{fig:GQDising}. 
This is quite a remarkable result, as it showed for the first time that GQD scales with universal critical exponents in the proximity of a quantum phase transition.

\begin{figure}[t]
\includegraphics[width=0.47\columnwidth]{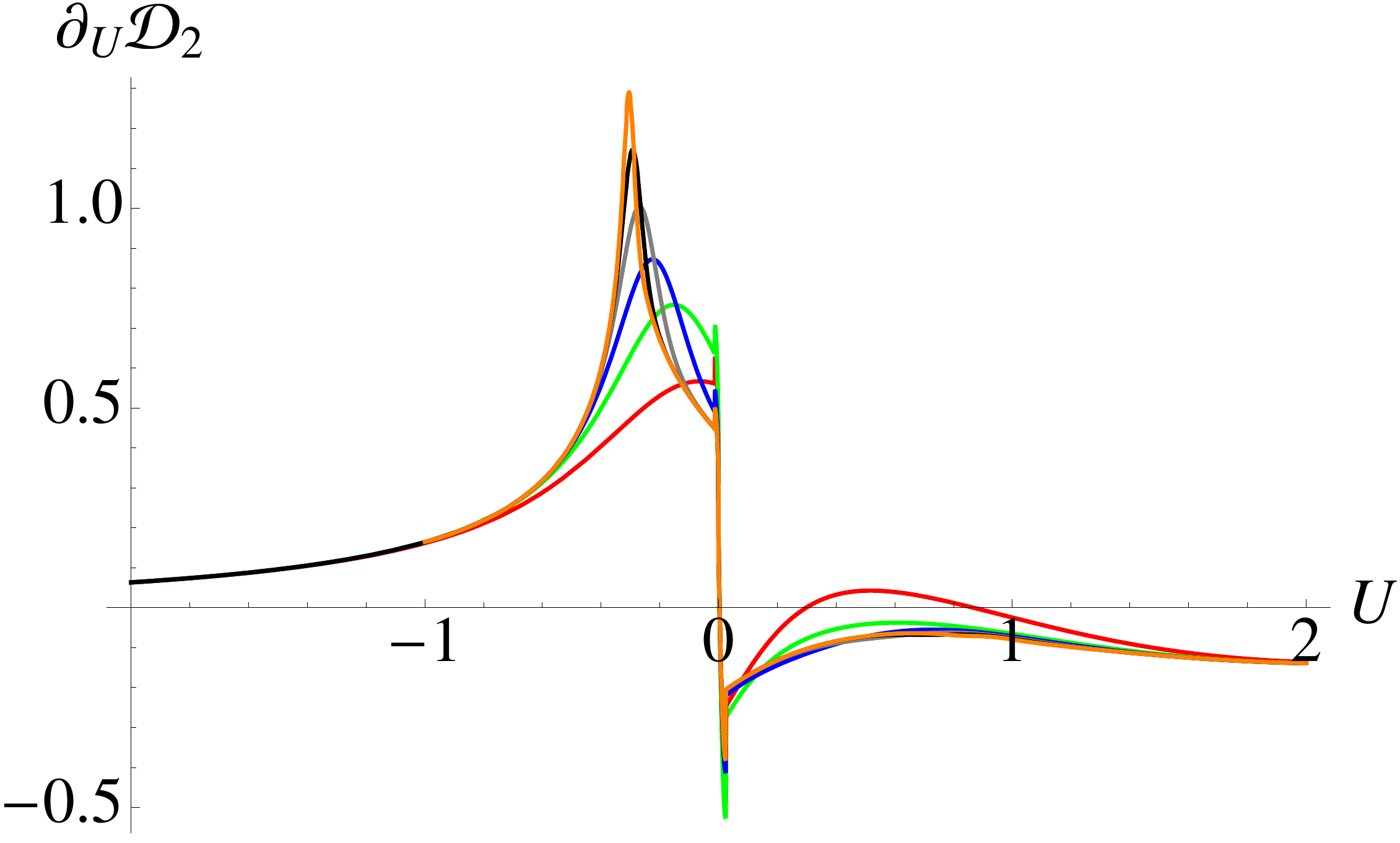}
\includegraphics[width=0.47\columnwidth]{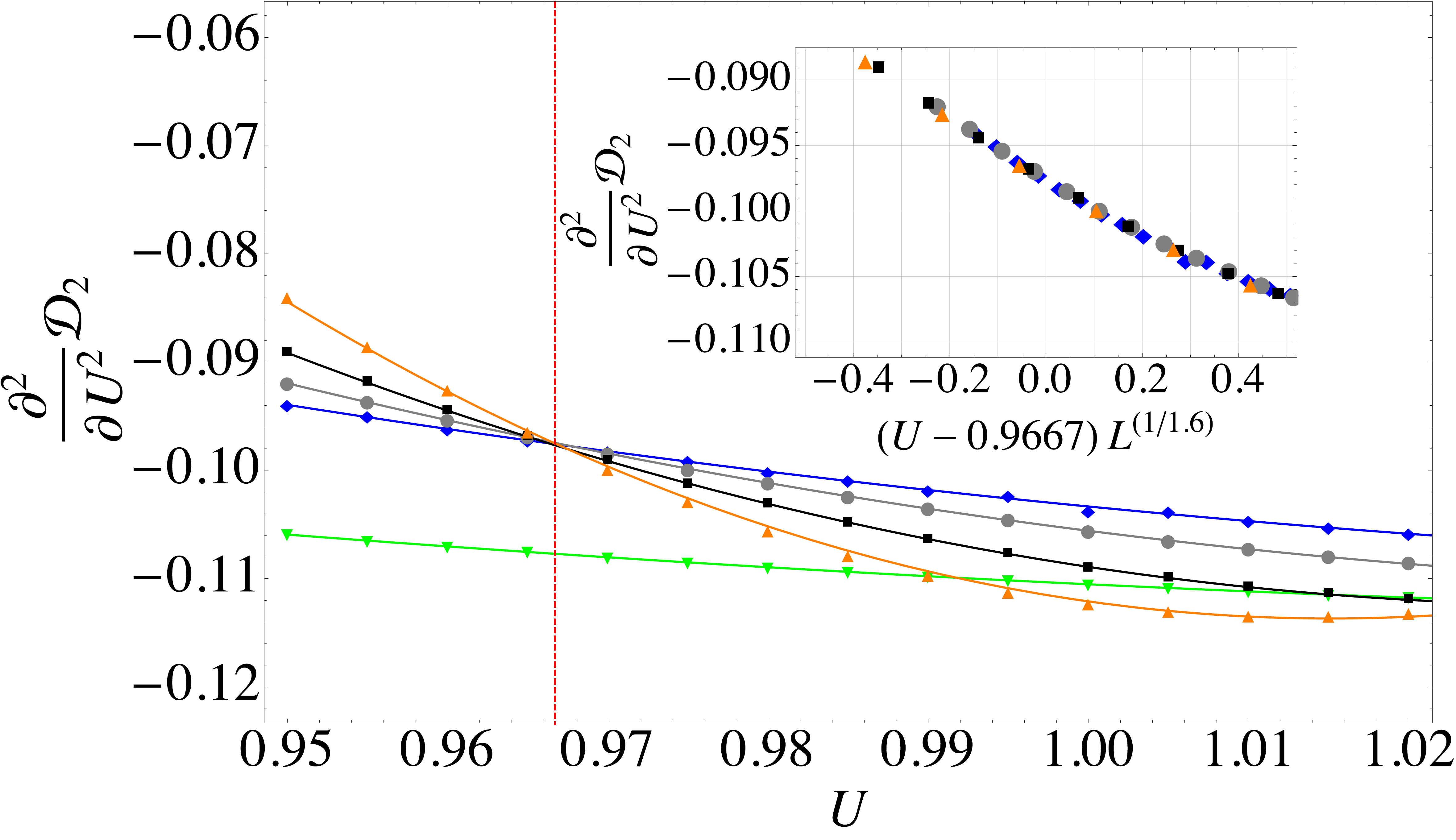}
\caption{(Color online) Left: First derivative of the nearest-neighbour symmetric discord against $U$ for $L=$ 8 (red), 16 (green), 32 (blue), 64 (gray), 128 (black) and 256 (orange). Right: Second derivative of $\mathcal{D}_2$ against $U$ in the vicinity of the Haldane-Large-D phase. 
Inset: collapse plot according to the ansatz Eq.~\eqref{eq:spin1ansatz}.
From Ref.~\cite{Power}.}
\label{fig:spin1GQD}
\end{figure}

Similar results have been obtained in \cite{Power} for the ground and thermal states of a spin-1 chain described by the Heisenberg Hamiltonian with uniaxial anisotropy of strength $U$ (see Eq.\eqref{eq:bilin}).
Despite the increased difficulty of performing the minimisation in Eqs.~\eqref{eq:discordsymm} and ~\eqref{GQD}, in Ref.~\cite{Power} the spin-1 case was analysed and $\mathcal D_2$  was obtained for two nearest-neighbour spins in the middle of a long chain by means of DMRG. Also $\mathcal{D}_N(\rho)$,  for small chains up to $N=6$ spins was obtained by means of exact diagonalisation. The behaviour of the symmetric bipartite discord is quite interesting as shown in Fig.~\ref{fig:spin1GQD}. In the left panel, the first derivative of $\mathcal{D}_2$ is plotted. In the vicinity of the second order N\'eel-Haldane transition, the first derivative correctly diverges with the system's length. Close to the third order Haldane-Large D transition instead, the first derivative has a point of inflection. Thus one expects the {\it second} derivative to follow a finite-size scaling ansatz:
\begin{equation}
\label{eq:spin1ansatz}
\frac{\partial^2\mathcal{D}_2}{\partial U^2} = f[(U-U_C)L^{1/\nu}],
\end{equation}
where, after fitting the results, leads to $U_C\simeq 0.9667$ and $\nu=1.6\pm0.1$ which are compatible with previous calculations \cite{Hu2011}. Thus in this case, as for the spin-1/2 case, quantum discord scales universally at second and third order quantum phase transitions. 

For thermal states, Power et al. \cite{Power} found that, generally, temperature destroys global discord but in some cases the opposite occur. 
We close this section by reporting an interesting development for the calculation of $\mathcal{D}_N(\rho)$ of a block of $N$ spins in translational invariant infinite chains by means of MPS techniques~ \cite{SunAnnalsPhysics2015,sun2014global}. There  authors found a more efficient way of storing an approximate version of the reduced density matrix of the $N$ spins that needs an amount of memory that scales polynomially with $N$. With this economical approach they find a recursive way of calculating ${\cal D}_N$ in terms of ${\cal D}_{N-1}$ and two-particle correlators that are easy to calculate with MPS techniques. They applied this method to the spin-1/2 XXZ model and for a three-body interaction model finding the scaling of  ${\cal D}_N$ close to the critical points.

\subsection{Quantum { coherence-based correlations} and Wigner-Yanase skew information}
The theory of quantum coherence has recently attracted much attention in the quantum information community and we refer the interested reader to a dedicated review~\cite{streltsov2016quantum}. Some quantum correlation features can be enhanced from the properties of quantum coherence of a state as a result of a quantum superposition in a certain basis.  

A measure of coherence with respect to a certain observable $K$ is embodied by the Wigner-Yanase skew information (WYSI) originally defined in \cite{wigner1963information}. The definition reads:
\begin{equation}
I(\rho,K)=-\frac 12 {\rm Tr}[\sqrt{\rho},K]^2
\end{equation}
From the definition it is clear that $I(\rho,K)$  accounts for the amount of non-commutativity (skewness) between the state $\rho$ and the observable $K$. In this sense it represents the uncertainty on the results of a measurement $K$ due to the coherence of $\rho$ in the basis of eigenstates of $K$. Girolami recently formally showed that the WYSI can indeed be used as a genuine form of quantum coherence for finite dimensional systems \cite{GirolamiPRL2014}. He also proposed a modified quantity $I^L(\rho,K) =-1/4{\rm Tr}[\rho,K]^2$ that, because of the absence of the square root of $\rho$, can be more easily computed and experimentally measured. Then he proved that the modified quantity sets a non-trivial lower bound on the WYSI: $I(\rho,K)\ge I^L(\rho,K)$.

For a bipartite $\rho_{AB}$ a measure of quantum correlations proposed in \cite{LuoFuOh} can be obtained from the WYSI of a local observable $K_A\otimes \mathbbm{1}_B$, in fact the difference:
\begin{equation}
F(\rho_{AB}) = I(\rho_{AB},K_A\otimes \mathbbm{1}_B) - I(\rho_{A},K_A) 
\end{equation}
quantifies the the amount of global quantum coherences relative to observable $K_A$ and vanishes for classically correlated states.

A similar quantity $I(\rho_{AB},K_A\otimes \mathbbm{1}_B)$ was employed by Karpat and coworkers to measure the local quantum coherence (LQC) \cite{CakmakPLA2012,Karpat2014,CakmakEntropy2015}. They analysed the scaling of the LQC in the 1D anisotropic XY chain for the reduced state of two nearest-neighbour spins and for $K_A=\sigma_x$. It was showed that the first derivative of the LQC diverges in proximity of the quantum phase transition. It is however worth reporting that occasionally these measures exhibit non-analytic behaviour, for example in the case of a factorisation point, even if the ground/thermal state of the chain is analytic. 

For the same model, it was shown in \cite{Lei2016} that the divergence with the system size of the first derivative of the LQC at the critical point is logarithmic. They also analyse spin chains with three-spin interactions finding practically no size-dependence on the local quantum coherence. Cheng et al. \cite{Cheng2015} showed a finite-size scaling ansatz for the first derivative of the local quantum coherence analogous to the one used for the two-qubit concurrence. Similar results have also been obtained for the extended Fermi-Hubbard model by means of DMRG calculations \cite{LiLin2016} and for topological quantum phase transitions in the 1D Kitaev chain and in the 2D Kitaev honeycomb model \cite{ChengEPLtopological}. The finite temperature scaling of the WYSI was analysed in \cite{Cheng2016} for infinite XY chains using a similar scaling ansatz which allows one to collapse  on the same curve data obtained for different temperatures.

More recent works have analysed the global quantum coherence, not directly related to quantum correlations,  in the ground and thermal states of one-dimensional spin chains. Quantum coherence has been defined in the context of a resource theory in \cite{Baumgratz,streltsov2016quantum}  
Mathematically it can be written as:
\begin{equation}
C(\rho) = S(\rho_{diag})-S(\rho)
\end{equation}
where $\rho_{diag}$ is obtained from $\rho$ by removing all the non-diagonal elements in a specific basis. Considering a family of states $\rho_\lambda$ defined in terms of a continuous parameter $\lambda$, Chen et al. defined the coherence susceptibility as $\partial_\lambda C(\rho_\lambda)$~\cite{CoherenceSuscept2016}. They showed the divergence of the coherence susceptibility for various examples of 1D chains including the Ising and XX models and for the 2D Kitaev honeycomb model.

\subsection{Mutual information}

\subsubsection{Quantum mutual information}
The popular measure of entanglement between two blocks of spins or lattice sites of a pure state by means of the von Neumann entropy, discussed in Sec.~\ref{sec:bipartite}, ceases to work when the overall state of the two blocks is not pure. This can happen if the two blocks are embedded in a larger lattice or when the system is at a non-zero temperature. 
For strongly correlated systems it has been proposed to use the mutual information which quantifies the total amount of correlations, classical and quantum, shared between two parties. The mutual information $I(\rho_{AB})$ is defined as:
\begin{equation}
I(\rho_{AB})=S(\rho_A)+S(\rho_B)-S(\rho_{AB})
\end{equation}
where $S(\rho)$ is the von Neumann entropy. It is easy to prove that the mutual information is non-negative and vanishes only for product states, $\rho_{AB}=\rho_A\otimes\rho_B$ as well as it is monotone under any completely positive trace preserving (TCPTP) map.
Because of the difficulty of evaluating the von Neumann entropy in quantum MonteCarlo methods, a modified Renyi mutual information has been defined:
\begin{equation}
I_n(\rho_{AB})=S_n(\rho_A)+S_n(\rho_B)-S_n(\rho_{AB})
\end{equation}
where $S_n(\rho) = (1-n)^{-1}\ln[{\rm Tr}(\rho^n)]$ is the Renyi entropy of order $n$. Notice that the Renyi entropy for $n>1$ can be negative and it therefore does not have the same operational meaning of the normal quantum mutual information. 

In contrast to the entanglement entropy of the ground state of gapped short-range Hamiltonians in 1D lattices, the mutual information fulfils an area law \cite{EisertRMP2010} regardless of the energy gap or the dimensionality of the lattice \cite{WolfPRL2008,Bernigau2015}. 
Most of the works on the quantum mutual information are concerned with ground and thermal states of one-dimensional lattice systems \cite{ZnidaricPRA2008,MelkoPRB2010,SinghPRL2011,UmPark,BonnesPRB2013}, two-dimensional arrays of spins \cite{SinghPRL2011} and in fully connected models as the Lipkin-Meshkov-Glick model \cite{Wilms}.
Hamma and coworkers recently proved that the quantum mutual information between macroscopically separated regions of spontaneous symmetry breaking ground state of one-dimensional spin chains vanish \cite{HammaPRA2016}. This does not just happen when the symmetry-breaking ground states are product states, as for ferromagnetic states and at factorisation points. 

The evolution of the quantum mutual information following a quench of the Hamiltonian has been described in Ref.~\cite{LauchliKollathJSTAT2008} for the 1D Bose-Hubbard model, in Refs.~\cite{SchachenmayerPRX2013,RegemortelPRA2016,LeporiJSTAT2017} for long-range Kitaev and Ising chains in connection to the Lieb-Robinson bound and the existence of a light cone for information propagation, and in Ref.~\cite{Nezhadhaghighi} for arrays of quantum harmonic oscillators with short and long-range interactions.
Violation of the area law has been predicted in the non-equilibrium steady state of a 1D fermionic lattice system when two halves at different temperatures are joined together \cite{EislerZimborasPRA2014}. The authors calculate the mutual information in the steady state of two adjacent blocks and find that it grows logarithmically with the block size. Similar results have been obtained for the spin-1/2 XY chain \cite{Ajisaka}. Recently, Kormos and Zimboras studied the dynamics of the transverse-field Ising chain after two chains at different temperatures are joined~\cite{kormos2016temperature}. Surprisingly, they found that the Renyi mutual information is in some instances {\it negative} and that it {\it decreases} with the block sizes. A systematic study of this phenomenon is still missing and there are many open questions.

For open quantum systems, the evolution of the quantum mutual information for the XY model in contact to a thermal reservoir made of a set of quantum harmonic oscillators \cite{Patane2009}. The authors find a crossover between a universal evolution for short times when the chain is brought close to its ground state critical point and a long-time equilibration dynamics. Very recently the entanglement and quantum mutual information for a fermionic quantum wire subject to two different temperatures at each end have been considered \cite{zanoci2016entanglement}. 

The quantum mutual information has also been extensively studied in the context of conformal field theory (CFT) where the scaling with the size of the block and their distance \cite{Furukawa,Nienhuis,CalabreseJSTAT2009,CardyJPA2013} and the time evolution following a quantum quench \cite{Asplund,Coser2014} have been analysed. See also Ref.~\cite{CalabreseCardyReview2009} for a dedicated review.

For random XX spin-chains the quantum mutual information of excited and thermal states has been studied in Ref.~\cite{HuangMoorePRB2014}.

We close this section by discussing an experimental measurement of the quantum mutual information in a two-dimensional optical lattice of bosonic atoms \cite{Kaufman2016}. Thanks to the single atom microscope, this experimental team was able to study the microscopic evolution of the atoms after a quench and measure the block Renyi entropies and the corresponding mutual information finding that they both follows a volume law.

\subsubsection{Shannon-Renyi mutual information}
In the last years a few works in condensed matter physics have employed a different form of mutual information based on the Shannon entropy. For a pure state expressed as a linear combination of the eigenstates $\ket i$ of an observable $O$:
\begin{equation}
\ket\psi = \sum_i c_i \ket i
\end{equation}
the Shannon entropy is defined as:
\begin{equation}
Sh(\psi) = -\sum_i p_i \log p_i
\end{equation}
where $p_i=|c_i|^2$ is the probability that when measuring the observable $O$ on state $\ket\psi$ one obtain the eigenvalue associated with $\ket i$. This can be easily employed for a mixed state $\rho$ for which $p_i=\bra i\rho\ket i$. The Shannon entropy has a well defined operational meaning in terms of the ignorance on the possible outcome of the measurement and can therefore be experimentally measured by observing the outcome statistics.  However it is basis dependent, similarly to other coherence measures. In analogy to Renyi entropy, one can measure the  Renyi-Shannon entropy: $Sh_n(\psi) =(1-n)^{-1}\ln \left(\sum_i p_i^n\right)$ and the corresponding Renyi-Shannon mutual information of a bipartite state $\rho_{AB}$: $I_{Sh,n}(\rho_{AB}) = Sh_n(\rho_{A})+Sh_n(\rho_{B})-Sh_n(\rho_{AB})$.

Renyi-Shannon entropies and mutual information have been studied for spin chains and for conformal field theories \cite{StephanPRB2009,StephanPRB2011,ZaletelPRL2011,AlcarazRajabpourPRL2013,AlcarazRajabpourPRB2014,Stephan2014,LuitzPRL2014,AlcarazRajabpourPRB2015}. The results for the scaling of Renyi-Shannon entropies and mutual information for blocks are not as coherent as for the von Neumann entropy and quantum mutual information. It has been found that the Renyi-Shannon entropy grows logarithmically with the block size of a critical spin chain with a prefactor proportional to the central charge, similarly to the von Neumann entropy. However this result holds only if the basis in which the Shannon entropy is computed is conformal, e.g. for a spin chain, the basis of eigenstates of $\sigma_z$ is conformal; in a different basis, the prefactor of the logarithmic growth has no direct relation with the central charge \cite{AlcarazRajabpourPRL2013,AlcarazRajabpourPRB2014}.

\subsubsection{Total quantum mutual information and many-body-localisation}
Renyi entropies and mutual information and all their variants account for classical and quantum correlations shared between two systems and can therefore be categorised as measures of bipartite correlations. But quantum systems made of many particles can instead be endowed with more complex multipartite correlations. Recently, a form of multipartite correlation, called total correlations was employed in the study of many-body localisation   \cite{Goold2015,Pietracaprina,DeTomasiPRL2017,campbell2016dynamics}. 
The total mutual information (also called in some places total correlations) are defined as follows~\cite{Modi2010,RMPDiscord}. Let us consider an N-partite system $\rho$ and let $\pi=\pi_1\otimes\pi_2\dots\otimes\pi_N$ be an N-partite product state. The total mutual  correlations are defined as the minimum relative entropy (see \eqref{eq:relative_entropy}) between the state $\rho$ and any product state $\pi$:
\begin{equation}
T(\rho) = \min_{\{\pi\}} S(\rho||\pi)
\end{equation}
 It has been shown that the unique state that minimises the relative entropy is the product state of all the reduced density matrices of $\rho_m$ for each party $m$: $\pi=\rho_1\otimes\rho_2\otimes\dots \rho_m$~\cite{Modi2010}. After some algebraic manipulation we therefore obtain:
 \begin{equation}
T(\rho)=\sum_{m=1}^N S(\rho_m)- S(\rho)
\end{equation}
Since for $N=2$ this is equivalent to the quantum mutual information, we can interpret $T(\rho)$ as a multipartite generalisation of the quantum mutual information. In contrast to the global discord which measures quantum multipartite correlations, the total mutual information measures quantum and classical multipartite correlations, hence the adjective {\it total}.

\begin{figure}[h]
\begin{center}
\includegraphics{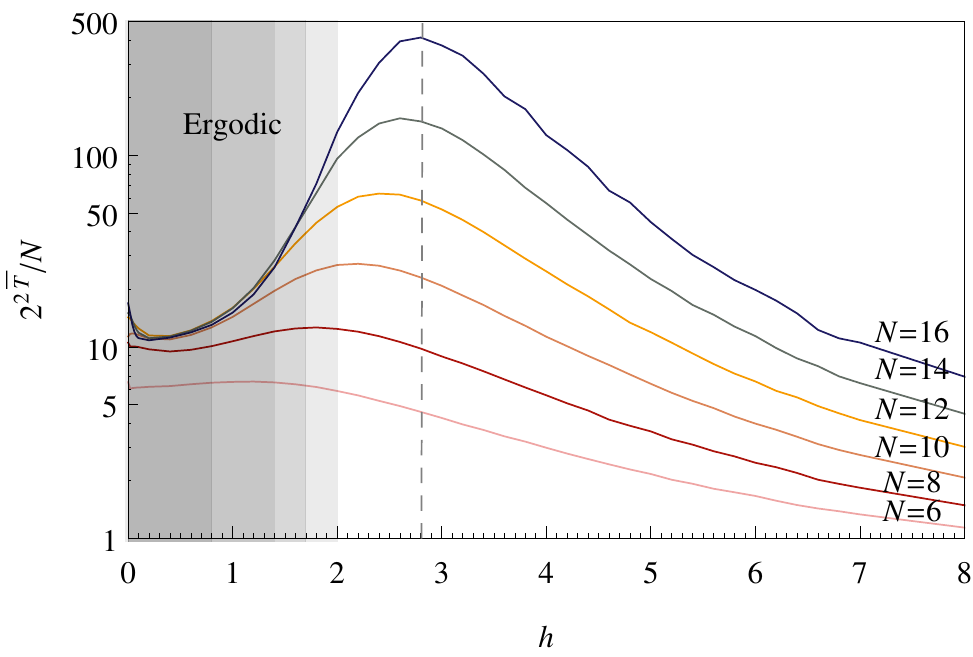}
\caption{Total correlations $T$ against the strength of the magnetic field disorder for different lengths $N=6,8,\dots,16$. (From Ref.~\cite{Goold2015})}
\label{fig:gooldmbl}
\end{center}
\end{figure}

In Ref.~\cite{Goold2015}, Goold and coworkers calculated $T(\rho)$ for a one-dimensional random chain that exhibits a dynamical phase transition between an ergodic phase and a many-body localised phase\footnote{As many-body localisation is a vast topic  we refer the reader to more specialised reviews~\cite{NandkishoreHuseReviewMBL,AltmanReviewMBL,LaflorencieReview,VasseurMooreReviewMBL,ImbrieRosReviewMBL}. Here we concentrate on the results of the total correlations.}
The model they considered is a Heisenberg spin chain with random magnetic fields:
\begin{equation}
H=\sum_{i=1}^N \left[J(\sigma_x^i\sigma_x^{i+1}+\sigma_y^i\sigma_y^{i+1}+\sigma_z^i\sigma_z^{i+1})+h_i\sigma_z^i\right ] 
\end{equation}
where $h_i$ are random magnetic fields uniformly distributed in the interval $[-h,h]$.
Goold and coworkers calculated the scaling of the total correlations, renormalised by the number of spins, of an ensemble of energy eigenstates in the middle of the energy spectrum, as a function of $h$. For small $h$, the system is in the ergodic phase and the total correlations are approximately proportional to $N$ indicating that strong correlations are established between all particles. For large $h$ instead, the system is many-body localised and the total correlations remain constant when increasing the system size. An example of the numerical results is shown in Fig.~\ref{fig:gooldmbl}, in which the extrapolation of the peak for infinite lengths occur at the ergodic-localised phase transition. 

Recently, in Ref.~\cite{campbell2016dynamics}, the time-dependent dynamics of the total correlations have been calculated for initial product states and product of singlets and compared to the behaviour of the nearest-neighbour concurrence. While the latter quantity strongly depends on the initial state, even within the class of product states, the total correlations are a more robust measure and can therefore distinguish more clearly the ergodic from the many-body localised phase.

\subsection{Metrology of strongly correlated systems}

\subsubsection{Entanglement and the quantum Fisher information}
In recent years a profound connection between the ultimate limit in the precision of parameter estimation in quantum systems and the presence of entanglement has been established \cite{TothPRA2012,LiLuo,HyllusPRA2012} (see also \cite{PezzeSmerziReview} for a pedagogical introduction).

Suppose a family of states, $\rho(\theta)$, depends on a phase $\theta$ generated by a Hermitian operator $O$: 
\begin{equation}
\rho(\theta) = e^{-i O\theta} \rho e^{iO\theta},
\end{equation}
  then it is possible to show that the uncertainty in the estimate of $\theta$ upon $m$ independent measurements is lower-bounded by the quantum Cramer-Rao bound:
\begin{equation}
\Delta\theta \ge \frac{1}{\sqrt{m F_Q}}
\end{equation}
where $F_Q$ is the quantum Fisher information (QFI) of $\rho(\theta)$:
\begin{equation}
F_Q(\rho(\theta),O) = {\rm Tr}[\rho(\theta)L^2_\theta]
\end{equation}
and the Hermitian operator $L_\theta$, called the symmetric logarithmic derivative, is the solution of the equation:
\begin{equation}
\frac{\partial \rho(\theta)}{\partial\theta} =\frac{\rho(\theta)L_\theta+L_\theta\rho(\theta)}{2}.
\end{equation}

For pure states, the QFI is simply the variance of the operator $O$: $F_Q=\bra\psi O^2\ket\psi-\bra\psi O\ket\psi^2$.

There has been a lot of debate on which states are useful for sensing and metrological applications and whether there is a connection with entanglement. In Refs.~\cite{TothPRA2012,LiLuo} this connection has been made stronger by a series of inequalities involving the quantum Fisher information and the presence of multipartite entanglement.  Let us consider a system of $N$ qubits and the collective angular momenta operators:
\begin{equation}
J_\alpha=\frac 12 \sum_i \sigma_\alpha^i
\end{equation}
where $\alpha=x,y,z$.
Then, for N-qubit separable states the following inequality holds \cite{PezzeSmerziPRL2009,TothPRA2012}:
\begin{equation}
\label{eq:fisherseparable}
F_Q(\rho_{sep},J_x)+F_Q(\rho_{sep},J_y)+F_Q(\rho_{sep},J_z)\le 2N
\end{equation}
thus any state violating Eq.~\eqref{eq:fisherseparable} must have some form of entanglement. Notice that the converse is in general true: there are entangled states that fulfil \eqref{eq:fisherseparable}.

Next, let us consider the concept of $k-$separability. A pure state is called $k-$producible if it is a tensor product of at most $k-$qubit states. A mixed state is  $k-$producible if it is a mixture of pure $k-$producible states. Thus a state that is not  $k-$producible is  $(k+1)-$multipartite entangled. For $N$ qubits,  the following inequality has been found for $k-$producible states:
 \begin{equation}
 \label{eq:Fisherk}
F_Q(\rho,J_\alpha)\le nk^2+(N-nk)^2
\end{equation}
where $n$ is the integer part of $N/k$.

These inequalities have been applied for spin systems close to a quantum phase transition in Refs.~\cite{MaWang,MaWang2013,NoriNJP2014,YePhysicaB}. In these papers it has been found that the QFI of the ground and thermal states of the system exhibit typical non-analyticities and scaling close to criticality. 

Recently, Heyl and coworkers \cite{hauke2016measuring} have shown that the QFI corresponding to a certain operator $O$ can be related to its dynamic susceptibility:
\begin{equation}
F_Q = \frac 4\pi \int_0^\infty d\omega \tanh\left(\frac{\omega}{2T}\right)\chi''(\omega,T)
\end{equation}
where $T$ is the temperature of the system and $\chi''(\omega,T)=\Im\chi(\omega,T)$ is the imaginary and dissipative part of the dynamic susceptibility:
\begin{equation}
\chi(\omega,T) = i\int_0^\infty dt e^{-i\omega t}{\rm Tr}(\rho [O(t),O])
\end{equation}
and $O(t)=e^{iHt}Oe^{-iHt}$. 

The great insight of this result is that the dynamic susceptibility of an operator, for example the total magnetisation $J_z$ of a spin chain, can in principle be measured in an experiment. In turn, violations of inequality \eqref{eq:Fisherk} can potentially demonstrate the presence of multipartite entanglement in equilibrium states of strongly correlated systems.

The multipartite entanglement of a Kitaev chain has been detected using the scaling quantum Fisher information~\cite{pezze2017multipartite} and put in relationship with the topological properties of the model.

Very recently, Pappalardi et al. \cite{pappalardi2017multipartite} studied the time evolution of the multipartite entanglement of a quantum state of a many-body system following a quantum quench. The multipartite entanglement is quantified through the QFI. The authors use as an example the quantum Ising chain whose transverse field is abruptly changed at the initial time. They find that the asymptotic state for long times contain more than two-partite entanglement, i.e. genuine multipartite entanglement.

\subsubsection{Metrology and thermometry}

When the parameter to be estimated is the temperature of an equilibrium state: 
\begin{equation}
\rho =\frac{ e^{-H/T} }{Z}, \quad Z={\rm Tr}e^{-H/T}
\end{equation}
the QFI associated to measuring the energy operator $H$ assumes a particularly simple form~ \cite{ZanardiParis2008,Invernizzi}:
\begin{equation}
\label{eq:Fisherthermal}
F_Q(H,\rho(T)) =\frac{ \Delta H^2}{T^4}
\end{equation}
and is thus related to the variance of the Hamiltonian and to the specific heat of the system. Thus, at a fixed temperature, the larger is the dispersion of the Hamiltonian, the more sensitive is the state to variations of the temperature; in turn this means that one can get a more accurate estimate of the temperature. This is true for relatively low temperatures while for high temperatures, the state becomes a mixture of many excited states and the sensitivity decreases due to the $T^4$ factor in the denominator of \eqref{eq:Fisherthermal}. On the other hand, if the sample is too cold, the above picture runs into troubles, especially when using individual quantum thermometers since the heat capacity of the probe drops down at low temperatures and correlations between the probe and the sample develop \cite{CorreaPRL2015,CorreaPRA2017}. 
In Ref.~\cite{ZanardiParis2008} it was discovered that quantum criticality of a many-body system can be used as a resource for estimating parameters of the Hamiltonian, like coupling parameters and magnetic fields, as well as the temperature. Intuitively, one can understand this result by noting that the state of the system near a quantum phase transition changes very quickly when one of the parameters of the Hamiltonian (or its temperature) is changed. 
In Ref.~\cite{Invernizzi} find an enhancement of the QFI proportional to the number of spins of an Ising chain at criticality. Other references discussing optimal parameter estimation include for spin chains Refs.~\cite{Skotiniotis2015, MehboudiPRA2016,BoyajianPRA2016,LiuSciRep2016} and for degenerate gases Ref.~\cite{MarzolinoPRA2013}.

Thermometry of strongly correlated systems is not an easy task. The theory explained above asserts that the optimal observable is the Hamiltonian of the system. However measuring the system energy of a quantum simulator realised with solid state or atomic systems is often unfeasible. In Ref.~\cite{MehboudiNJP2015}, Mehboudi et al. proposed a scheme to estimate the temperature of a spin chain realised with ultracold atoms by measuring collective operators. These operators, for example the total or staggered magnetisation in a certain direction are very good temperature estimators, and the authors have shown that the QFI associated with them can, in an intermediate range of temperatures be very close to the ultimate limit imposed by the quantum Cramer-Rao bound. The measurement of such collective operators could be realised by coupling the atoms to the field of a propagating laser or of an optical cavity.  Mehboudi et al. proposed to use the quantum Faraday rotation of the polarisation of an incoming laser beam induced by the atomic collective angular momentum as originally proposed in Ref.~\cite{Eckert2008}. One of the advantages of this method  is that the measurement is quantum non demolishing and could potentially be used to monitor the system temperature in time without destroying the sample. 

The previous approach was based on the measurement of collective operators thus giving access to the temperature of the system as whole. But what is the precision limit of an estimate of the temperature based on local measurements? This could be particularly useful when the thermometry is a local probe like an impurity interacting locally with a many-body system. De Pasquale et al.~\cite{DePasquale2016}, defined a local quantum thermal susceptibility (LQTS), a microscopic quantity that reduces to the heat capacity for macroscopic thermometers, that is able to locally distinguish the ground state and the first excited subspace. Therefore it is a very precise thermometer in the regime of very small temperatures in which quantum phase transitions occur. The authors showed that the sensitivity of such a thermometer is able to detect the phase transitions occurring for the quantum Ising and XXZ chains.

\subsubsection{Metrology and dissipative quantum phase transitions}
We close this section on metrology of strongly correlated systems by discussing recent developments in the are of open quantum systems. A new class of dynamical phase transitions \cite{GarrahanLesanovskyPRL2010} has been introduced to describe the dramatic change of dynamical behaviour in the steady state of an open quantum system. It has been recently argued that the estimation of a parameter of an open quantum system can be enhanced when the system is close to such a dynamical phase transition~\cite{Macieszczak}. In particular, the QFI scales quadratically with the observation time of the system near a first-order dynamical phase transition instead of the typical linear scaling. Other investigations of the QFI in the nonequilibrium steady states of quantum spin chains include Refs.~\cite{MarzolinoPRA2014,MarzolinoPRB2017}.

%


%% file: Chapter6.tex

\section{Discussion}

\label{sec:conclussions}
In this review we have summarised the latest development regarding quantum correlations in quantum many-body systems. The role of quantum correlations to understand, classify and simulate quantum many body systems has become unavoidable in order to bring a new and unified picture of quantum many-body systems which goes beyond 
the conventional paradigm of band theory, conventional superconductivity (BCS theory) and  the Ginzburg-Landau theory of phase transitions. Traditionally, the characterisation of quantum phases and their transitions had been based mostly on local order parameters and the response to linear perturbations given by low order correlators, e.g. spin-spin correlates $\langle S_{i}S_{j}\rangle$. Those have been very useful to explain order associated to the breaking of some symmetry as formalised in the Ginzburg-Landau theory of phase transitions.  However,  the frontiers of quantum matter have broadened substantially with the presence of topological  and other exotic states of matter. 

The theory of quantum correlations, especially entanglement, is key in the continuous development of efficient techniques for simulating classically quantum many-body systems. Currently, methods borrowed from machine learning are being applied for diagnostic of quantum many-body states \cite{Carleo602} and can be useful in the analysis of their quantum correlations~\cite{DengPRX2017,deng2017machine}. 

  Such theoretical developments run in parallel and
are stimulated by the spectacular experimental progress
achieved in the areas of ultracold atomic physics and condensed
matter. New platforms such as ion traps, optical lattices,
bosonic and fermionic atomic degenerate gases, new
superconducting materials, quantum Hall systems
are intensively investigated as quantum simulators. On
these devices a huge set of new phases is postulated to
be synthesisable. Besides, the quest for the creation of exotic materials in solid state systems exhibiting quantum features such as spin liquids, and topological insulators, the quantum simulation of quantum many-body models with controlled quantum systems, following Feynman's idea, is currently opening the exploration of new quantum phenomena with unprecedented control. 

There remain many open problems in this area. While for 1D systems the characterisation of different phases and their transitions is well understood in terms of entanglement and other quantum correlations, the situation for 2D system is much less clear. Moreover, the recent advances in the understanding of quantum steering, non signalling theory  and other generalised resource theories could result in new diagnostic tools for a more complete characterisation of quantum many-body systems.

%
%